\documentclass[a4paper,11pt]{article}

\usepackage{fullpage}
\usepackage{comment}

\usepackage{latexsym}
\usepackage{amsthm}
\usepackage{amsmath}
\usepackage{amssymb}
\usepackage{graphicx}
\usepackage{graphics}
\usepackage{xspace}

\usepackage{tikz}
\usetikzlibrary{calc}
\usetikzlibrary{shapes}

\newcommand{\vca}{\index{vertex cover!number}Vertex Cover\xspace}
\newcommand{\mla}{\index{max leaf!number}Max Leaf\xspace}
\newcommand{\lfa}{\index{linear forest!distance to}Distance to Linear Forest\xspace}

\newcommand{\cliquea}{\index{clique!distance to}Distance to a Clique\xspace}
\newcommand{\fvsa}{\index{feedback vertex number}\index{forest!distance to}Feedback Vertex Set\xspace}

\newcommand{\chordala}{\index{graph!chordal!distance to}Distance to Chordal\xspace}
\newcommand{\intervala}{\index{graph!interval!distance to}Distance to Interval\xspace}
\newcommand{\opa}{\index{graph!outerplanar!distance to}Distance to Outerplanar\xspace}
\newcommand{\pwa}{\index{pathwidth}Pathwidth\xspace}

\newcommand{\clustera}{\index{graph!cluster!distance to}Distance to Cluster\xspace}
\newcommand{\coclustera}{\index{graph!cocluster!distance to}Distance to Cocluster\xspace}
\newcommand{\cographa}{\index{graph!cograph!distance to}Distance to Cograph\xspace}

\newcommand{\octa}{\index{odd cycle transversal!number}Odd Cycle Transversal\xspace}

\newcommand{\twa}{\index{treewidth}Treewidth\xspace}
\newcommand{\perfecta}{\index{graph!perfect!distance to}Distance to Perfect\xspace}

\tikzset{node/.style={
circle,
very thick,
draw=black!90!white!90, % 50% red and 50% black,
top color=white, % a shading that is white at the top...
bottom color=red!50!black!20, % and something else at the bottom
inner sep=0,
minimum size=0.5cm,
align=center
}}

\tikzset{param/.style={
rounded rectangle,
minimum height=0.3cm,
very thick,
draw=black!90!white!90, % 50% red and 50% black,
top color=white, % a shading that is white at the top...
bottom color=red!50!black!20, % and something else at the bottom
text width=2cm,
rounded corners=2pt,
text centered,
thin
}}

\tikzset{legend/.style={
rectangle,
minimum width=0.8cm,
minimum height=0.2cm,
very thick,
draw=black!90!white!90, % 50% red and 50% black,
top color=white, % a shading that is white at the top...
bottom color=red!50!black!20, % and something else at the bottom
rounded corners=2pt,
rounded corners=2pt,
thin
}}

\tikzset{paraNPC/.style={
very thick,
dotted,
top color=white,
bottom color=black!50!white,
}}

\tikzset{FPT/.style={
double=white,
double distance=1pt,
top color=white,
bottom color=black!10!white,
}}

\tikzset{WhardInXP/.style={
top color=white,
bottom color=black!20!white,
dashed
}}

\tikzset{UnknownComplexity/.style={
top color=white,
bottom color=black!20!white,
dashed
}}

\tikzset{InXP/.style={
double=white,
double distance=1pt,
top color=white,
bottom color=black!30!white,
}}

\tikzset{fptNoKernels/.style={
bottom color=black!30!red!60, top color=white,
very thick
}}

\tikzset{fptWithKernels/.style={
top color=green!30, bottom color=white
}}

\tikzset{edgeBound/.style={
midway,
text centered,
sloped
}}

\tikzset{edgeBoundHor/.style={
midway,
text centered,
}}
\newcommand{\treewidthHierarchyPicture}{
\begin{figure}[t]
	\small

	\centering
\begin{tikzpicture}[inner sep=0.1cm, >=stealth, thick, line cap=round]

\coordinate (dx) at (4cm,0cm);
\coordinate (dy) at (0cm,-1.5cm);

\definecolor{c1}{RGB}{235,235,235}

\node (vc) at ($0.75*(dx) + -1*(dy)$) [param,fptWithKernels] {\vca \S\ref{sect:tw:byVC}};
\node (ml) at ($1.5*(dx) + 0*(dy)$) [param,fptWithKernels] {\mla \cite{KleitmanW91}};
\node (cluster) at ($0.75*(dx) + 1*(dy)$) [param,fptNoKernels] {\clustera \cite{BodlaenderJK11a}};
\node (clique) at ($0.5*0.75*(dx) + 0*(dy)$) [param,fptNoKernels] {\cliquea \S\ref{sect:tw:fptClassifications}};
\node (cocluster) at ($0*(dx) + 1*(dy)$) [param,fptNoKernels] {\coclustera \cite{BodlaenderJK11a}};
\node (cograph) at ($0*(dx) + 2*(dy)$) [param,UnknownComplexity] {\cographa};
\node (lf) at ($1.5*(dx) + 1*(dy)$) [param,fptWithKernels] {\lfa};
\node (fvs) at ($1.5*(dx) + 2*(dy)$) [param,fptWithKernels] {\fvsa \S\ref{sect:tw:byFVS}};
\node (interval) at ($0.75*(dx) + 2*(dy)$) [param,UnknownComplexity] {\intervala};
\node (chordal) at ($0.5*0.75*(dx) + 3*(dy)$) [param,UnknownComplexity] {\chordala};
\node (op) at ($0.5*3.75*(dx) + 3*(dy)$) [param,FPT] {\opa};
\node (pw) at ($2.25*(dx) + 2*(dy)$) [param,fptNoKernels] {\pwa};
\node (oct) at ($0.5*2.25*(dx) + 3*(dy)$) [param,paraNPC] {\octa \cite{KloksK95}};
\node (tw) at ($2.25*(dx) + 4*(dy)$) [param,fptNoKernels,text width=2.4cm] {\twa \cite{Bodlaender96}};
\node (perfect) at ($0*(dx) + 4*(dy)$) [param,paraNPC] {\perfecta};

\draw[out=-60,in=150] (vc) to (lf);
\draw (vc) -- (cluster);
\draw[out=180,in=100,shorten >=1pt] (vc) to (cocluster);
\draw[shorten >=1pt] (cluster) -- (cograph);
\draw[shorten >=1pt,shorten <=1pt] (cocluster) -- (cograph);
\draw[in=-110,out=110,shorten >=1pt] (perfect) to (cograph);
\draw (clique) -- (cluster);
\draw (clique) -- (cocluster);
\draw (cluster) -- (interval);
\draw (ml) -- (lf);
\draw (lf) -- (fvs);
\draw (lf) to (pw);
\draw (lf) -- (interval);
\draw (interval) -- (chordal);
\draw[out=200,in=25] (fvs) to (chordal);
\draw[shorten >=1pt] (fvs) -- (op);
\draw (fvs) -- (oct);
\draw (chordal) -- (perfect);
\draw[shorten <=1pt] (op) -- (tw);
\draw[out=-70,in=70] (pw) to (tw);
\draw (oct) -- (perfect);

\end{tikzpicture}

	\caption[Complexity overview for various parameterizations of \treewidth.]{Complexity overview for various parameterizations of \treewidth. Larger parameters are drawn higher. For deletion distance parameters we assume that a deletion set is given along with the input. The shading indicates that a parameterization is either 
	FPT with currently unknown kernelization complexity
	\protect\tikz{\protect \node at (0,0) [legend,FPT] {};},
	para-NP-complete 
	\protect\tikz{\protect \node at (0,0) [legend,paraNPC] {};},
	FPT but (conditionally) lacking a polynomial kernel
	\protect\tikz{\protect \node at (0,0) [legend,fptNoKernels] {};},
	FPT with a polynomial kernel
	 \protect\tikz{\protect \node at (0,0) [legend,fptWithKernels] {};},
	or of unknown complexity
	 \protect\tikz{\protect \node at (0,0) [legend,UnknownComplexity] {};}.
	 Note that negative results are inherited by lower parameters and positive results by higher parameters~\cite{FellowsJR12}, e.g., the existence of polynomial kernels for the listed parameters is open only for distance to outerplanar.
	}
	\label{figure:treewidthHierarchyPicture}
\end{figure}
}
\newtheorem{observation}{Observation}
\newtheorem{theorem}{Theorem}
\newtheorem{lemma}{Lemma}
\newtheorem{proposition}{Proposition}
\newtheorem{corollary}{Corollary}
\newtheorem{claim}{Claim}

\theoremstyle{definition}
\newtheorem{definition}{Definition}

\newtheorem{rul}{Rule}[section]

\newcommand{\nolabel}[1]{}

\newcommand{\eweight}{\ensuremath{\mathop{\mathrm{\textsc{E-weight}}}}\xspace}
\newcommand{\F}{\ensuremath{\mathcal{F}}\xspace}
\newcommand{\X}{\ensuremath{\mathcal{X}}\xspace}

\newcommand{\R}{\ensuremath{\mathcal{R}}\xspace}
\newcommand{\T}{\ensuremath{\mathcal{T}}\xspace}
\newcommand{\C}{\ensuremath{\mathcal{C}}\xspace}
\newcommand{\low}[0]{\ensuremath{\mathrm{\textsc{low}}}\xspace}

\newcommand{\Oh}{\mathcal{O}}

\newcommand{\tw}{\textsc{tw}}
\newcommand{\yes}{\textbf{yes}\xspace}
\newcommand{\no}{\textbf{no}\xspace}
\newcommand{\vc}{\ensuremath{\mathrm{\textsc{vc}}}\xspace}
\newcommand{\fvs}{\ensuremath{\mathrm{\textsc{fvs}}}\xspace}
\newcommand{\containment}[0]{NP~$\subseteq$~coNP$/$poly\xspace}
\newcommand{\notcontainment}[0]{NP~$\not \subseteq$~coNP$/$poly\xspace}

\newcommand{\treewidth}[0]{\textsc{Treewidth}\xspace}
\newcommand{\witreewidth}[0]{\textsc{Weighted Independent Treewidth}\xspace}
\newcommand{\pathwidth}[0]{\textsc{Pathwidth}\xspace}
\newcommand{\feedbackvertexset}[0]{\textsc{Feedback Vertex Set}\xspace}
\newcommand{\weightedtreewidth}[0]{\textsc{Weighted Treewidth}\xspace}
\newcommand{\treewidthByChordal}[0]{\textsc{Treewidth parameterized by a modulator to chordal graphs}\xspace}
\newcommand{\treewidthByCograph}[0]{\textsc{Treewidth parameterized by a modulator to a cograph}\xspace}
\newcommand{\cutwidththree}[0]{\textsc{Cutwidth3}\xspace}
\newcommand{\cutwidth}[0]{\textsc{Cutwidth}\xspace}
\newcommand{\TWMSC}[0]{\textsc{tw by clique}\xspace}
\newcommand{\treewidthByVC}[0]{\textsc{Treewidth parameterized by a vertex cover}\xspace}
\newcommand{\treewidthByFVS}[0]{\textsc{Treewidth parameterized by a feedback vertex set}\xspace}
\newcommand{\pathwidthByVC}[0]{\textsc{Pathwidth parameterized by a vertex cover}\xspace}
\newcommand{\pathwidthByFVS}[0]{\textsc{Pathwidth parameterized by a feedback vertex set}\xspace}
\newcommand{\treewidthByClique}[0]{\textsc{Treewidth parameterized by a modulator to a single clique}\xspace}
\newcommand{\pathwidthByClique}[0]{\textsc{Pathwidth parameterized by a modulator to a single clique}\xspace}
\newcommand{\weightedTreewidthByVC}[0]{\textsc{Weighted Treewidth parameterized by a vertex cover}\xspace}
\newcommand{\weightedTreewidthByClique}[0]{\textsc{Weighted Treewidth parameterized by a modulator to a single clique}\xspace}

\newcommand{\HamCycle}[0]{\textsc{Hamiltonian Cycle}\xspace}
\newcommand{\ThreeColoring}[0]{\textsc{$3$-Coloring}\xspace}
\newcommand{\oddcycletransversal}[0]{\textsc{Odd Cycle Transversal}\xspace}

\newcommand{\problemdef}[3]
{
\begin{quote}
\textsc{#1}\\
\textbf{Instance:} #2\\
\textbf{Question:} #3
\end{quote}
}

\newcommand{\parproblemdef}[4]
{
\begin{quote}
\textsc{#1}\\
\textbf{Instance:} #2\\
\textbf{Parameter:} #3\\
\textbf{Question:} #4
\end{quote}
}

\newcommand{\chaptref}[1]{\hyperref[#1]{Chapter~\ref{#1}}}
\newcommand{\caseref}[1]{\hyperref[#1]{Case~\ref{#1}}}
\newcommand{\casesref}[1]{\hyperref[#1]{Cases~\ref{#1}}}
\newcommand{\partref}[1]{\hyperref[#1]{Part~\ref{#1}}}
\newcommand{\sectref}[1]{\hyperref[#1]{Section~\ref{#1}}}
\newcommand{\ssectref}[1]{\hyperref[#1]{Sect.~\ref{#1}}}
\newcommand{\subsectref}[1]{\hyperref[#1]{Subsection~\ref{#1}}}
\newcommand{\defref}[1]{\hyperref[#1]{Definition~\ref{#1}}}
\newcommand{\invariantref}[1]{\hyperref[#1]{Invariant~\ref{#1}}}
\newcommand{\lemmaref}[1]{\hyperref[#1]{Lemma~\ref{#1}}}
\newcommand{\lemmataref}[1]{\hyperref[#1]{Lemmata~\ref{#1}}}
\newcommand{\thmref}[1]{\hyperref[#1]{Theorem~\ref{#1}}}
\newcommand{\corollaryref}[1]{\hyperref[#1]{Corollary~\ref{#1}}}
\newcommand{\obsref}[1]{\hyperref[#1]{Observation~\ref{#1}}}
\newcommand{\tableref}[1]{\hyperref[#1]{Table~\ref{#1}}}
\newcommand{\tablesref}[1]{\hyperref[#1]{Tables~\ref{#1}}}
\newcommand{\stepref}[1]{\hyperref[#1]{Step~\ref{#1}}}
\newcommand{\imgref}[1]{\hyperref[#1]{Fig.~\ref{#1}}}
\newcommand{\figref}[1]{\hyperref[#1]{Fig.~\ref{#1}}}
\newcommand{\algref}[1]{\hyperref[#1]{Algorithm~\ref{#1}}}
\newcommand{\ruleref}[1]{\hyperref[#1]{Rule~\ref{#1}}}
\newcommand{\rulesref}[1]{\hyperref[#1]{Rules~\ref{#1}}}
\newcommand{\propref}[1]{\hyperref[#1]{Property~\ref{#1}}}
\newcommand{\proposref}[1]{\hyperref[#1]{Proposition~\ref{#1}}}
\newcommand{\opref}[1]{\hyperref[#1]{Operation~\ref{#1}}}
\newcommand{\exampleref}[1]{\hyperref[#1]{Example~\ref{#1}}}
\newcommand{\apref}[1]{\hyperref[#1]{Appendix~\ref{#1}}}
\newcommand{\claimref}[1]{\hyperref[#1]{Claim~\ref{#1}}}

\usepackage{xcolor}
\definecolor{dark-red}{rgb}{0.4,0.15,0.15}
\definecolor{dark-blue}{rgb}{0.15,0.15,0.4}
\definecolor{medium-blue}{rgb}{0,0,0.5}
\usepackage[hypertexnames=false]{hyperref}

\hypersetup{
    colorlinks, linkcolor={dark-red},
    citecolor={dark-blue}, urlcolor={medium-blue}
}

\begin{document}
\title{Preprocessing for Treewidth:\\ A Combinatorial Analysis through Kernelization\thanks{
This work was supported by the Netherlands Organization for Scientific Research (N.W.O.), project ``KERNELS: Combinatorial Analysis of Data Reduction''. An extended abstract of this work appeared in the Proceedings of the 38th International Colloquium on Automata, Languages and Programming~\cite{BodlaenderJK11b}. Our presentation here contains an improvement to the extended abstract, which appeared in the Proceedings of the 13th Scandinavian Symposium and Workshops on Algorithm Theory~\cite{BodlaenderJK12a}.}}
\author{Hans L.~Bodlaender\\Utrecht University\\The Netherlands\\\texttt{hansb@cs.uu.nl}
\and
Bart M. P. Jansen\\Utrecht University\\The Netherlands\\\texttt{bart@cs.uu.nl}
\and
Stefan Kratsch\\Technical University Berlin\\Germany\\\texttt{stefan.kratsch@tu-berlin.de}
}
% \institute{
% Department of Information and Computing Sciences,
% Utrecht University, P.O. Box 80.089, 3508 TB Utrecht, The Netherlands,
% {\tt $\{$hansb,bart,kratsch$\}$@cs.uu.nl}}
% \institute{
% Utrecht University, The Netherlands,
% {\tt $\{$hansb,bart,kratsch$\}$@cs.uu.nl}}

\maketitle

\begin{abstract}
The notion of treewidth plays an important role in theoretical and practical studies of graph problems. It has been recognized that, especially in practical environments, when computing the treewidth of a graph it is invaluable to first apply an array of preprocessing rules that simplify and shrink it. This work seeks to prove rigorous performance guarantees for such preprocessing rules, both known and new ones, by studying them in the framework of kernelization from parameterized complexity.

%It is known that the NP-hard problem of determining whether a given graph~$G$ has treewidth at most~$k$ admits no preprocessing algorithm that reduces any input instance to size polynomial in~$k$, unless NP~$\subseteq$~coNP/poly and the polynomial hierarchy collapses to its third level. In this paper, we consider different parameterizations of treewidth. We show that treewidth has a kernel with~$\Oh(\ell^3)$ vertices, where~$\ell$ denotes the size of a vertex cover, and a kernel with~$\Oh(\ell^4)$ vertices, where~$\ell$ denotes the size of a feedback vertex set. This implies that given an instance~$(G,k)$ of treewidth we can efficiently reduce its size to~$\Oh((\fvs(G))^4)$ vertices, where~$\fvs(G)$ is the size of a minimum feedback vertex set in~$G$. In contrast, we show that treewidth parameterized by the vertex-deletion distance to a single clique and weighted treewidth parameterized by the size of a vertex cover do not have polynomial kernels unless NP~$\subseteq$~coNP/poly.
It is known that the NP-complete problem of determining whether a given graph~$G$ has treewidth at most~$k$ admits no polynomial-time preprocessing algorithm that reduces any input instance to size polynomial in~$k$, unless NP~$\subseteq$~coNP/poly and the polynomial hierarchy collapses to its third level. In this paper we therefore consider structural graph measures larger than treewidth, and determine whether efficient preprocessing can shrink the instance size to a polynomial in such a parameter value.

We prove that given an instance~$(G,k)$ of treewidth we can efficiently reduce its size to~$\Oh(\fvs(G)^4)$ vertices, where~$\fvs(G)$ is the size of a minimum feedback vertex set in~$G$. We can also prove a size reduction to~$\Oh(\vc(G)^3)$ vertices, where~$\vc(G)$ is the size of a minimum vertex cover. Phrased in the language of parameterized complexity, we show that \treewidth has a polynomial kernel when parameterized by the size of a given feedback vertex set, and also by the size of a vertex cover. In contrast we show that \treewidth parameterized by the vertex-deletion distance to a single clique, and \weightedtreewidth parameterized by the size of a vertex cover, do not admit polynomial kernelizations unless \containment.
\end{abstract}

\section{Introduction}
\label{section:introduction}
\emph{Treewidth} is a well-studied graph parameter, with many theoretical and practical applications. A related parameter is \emph{Weighted Treewidth}, where vertices have weights and the width of a decomposition is given by the total weight of the vertices in a heaviest bag, minus one. In this work we study the decision problems related to these width parameters, which given a (weighted) graph~$G$ and integer~$k$ ask whether the (weighted) treewidth of~$G$ is at most~$k$. Precise definitions of these concepts are given in \sectref{section:definitions}.

Preprocessing heuristics for \treewidth and \weightedtreewidth have been studied in a practical setting \cite{BodlaenderK06c,BodlaenderKE05,EijkhofBK07}. The experimental results reported in these papers show that there are simplification routines that give significant size reductions for many practical instances, making it more feasible to compute the treewidth of those graphs exactly or approximately. However, these heuristics do not give any guarantees on the effectiveness of the preprocessing: there is no provable bound on the size of the processed instances. The purpose of this work is to give a theoretical analysis of the potential of preprocessing for \treewidth, studying whether there are efficient preprocessing procedures whose effectiveness can be proven, and what the resulting size bounds look like. Such investigations are made possible using the concept of \emph{kernelization}~\cite{Bodlaender09,GuoN07}, which is a relatively young subfield of algorithm design and analysis based on parameterized complexity theory~\cite{DowneyF99,FlumG06}. A \emph{kernelization algorithm} (or \emph{kernel}) is a polynomial-time algorithm which given an instance~$(x,k) \in \Sigma^* \times \mathbb{N}$ of some parameterized problem, computes an equivalent 
instance~$(x', k')$ whose size is bounded by a function~$f(k)$ depending only on the chosen parameter, i.e.,~$|x'|, k' \leq f(k)$. The function~$f$ is the \emph{size} of the kernel, and \emph{polynomial kernels} ($f \in k^{\Oh(1)}$) are of particular interest.

From a theoretical point of view, the fact that \treewidth is fixed-parameter tractable (FPT; see for instance \cite{Bodlaender96,LagergrenA91}), implies that there is a kernel for the problem. However, the size of such a kernel depends on the function of the parameter in the running time of the FPT algorithm; with the current state of FPT algorithms for \treewidth this size would be exponential in~$k^3$ (where~$k$ is the target treewidth). Bodlaender et al.~\cite{BodlaenderDFH09} have shown that \treewidth with standard parameterization (i.e., parameterized by~$k$) has no polynomial kernel unless all coNP-complete problems have distillation algorithms; they conjectured that the latter is impossible (the \emph{AND-distillation conjecture}~\cite{BodlaenderDFH09}). A recent breakthrough work of Drucker~\cite{Drucker12_focs} proved this conjecture under the assumption that \notcontainment. Hence it is unlikely that there is a polynomial-time algorithm that reduces the size of an instance~$(G,k)$ of \treewidth to a polynomial in the desired treewidth~$k$. We therefore turn to other parameters (e.g., the vertex cover number of the input graph), and determine whether we can efficiently shrink an input of \treewidth to a size which is polynomial in such a parameter. We consider different structural parameters of the input graph: these parameters measure the number of vertex deletions needed to transform the input into a member of some very simple graph class. All parameterized problems we consider fit the following template, where~\F is a class of graphs:
\parproblemdef{Treewidth parameterized by a modulator to~$\F$}
{A graph~$G$, a positive integer~$k$, and a set~$S\subseteq V(G)$ such that~$G-S\in\F$.}
{$\ell:=|S|$.}
{Is the treewidth of~$G$ at most~$k$?} %$\tw(G)\leq k$?}
The set~$S$ is a \emph{modulator} to the class \F. The problem \weightedTreewidthByVC is defined analogously, using the fact that a vertex cover is a modulator to an edgeless graph. We assume that a modulator is given along with the input to decouple the complexity of \emph{finding} a modulator from the complexity of \emph{exploiting its structural guarantees}. For a thorough discussion of the technicalities of structural parameterizations we refer to the survey by Fellows et al.~\cite{FellowsJR12}.

\paragraph{Our work.} In this paper, we add positive theoretical results to the positive experimental work. We first take the size of a vertex cover of~$G$ as the parameter, resulting in the problem \treewidthByVC (which fits into the given template when using~$\F$ as the class of edgeless graphs). We prove that this problem admits a polynomial kernel with~$\Oh(\ell^3)$ vertices. Since we can first compute a~$2$-approximation for the minimum vertex cover and then feed it to our kernelization algorithm, this implies that an instance~$(G,k)$ of \treewidth on a graph with a minimum vertex cover of size~$\vc(G)$ can be shrunk in polynomial-time into an instance with~$\Oh(\vc(G)^3)$ vertices, even if we are not given a minimum vertex cover in the input. 

We then turn to the parameter ``feedback vertex number'', which is easily seen to be at most the value of the vertex cover number. We extend our positive results by showing that \treewidthByFVS (which fits the template when~$\F$ is the class of forests) admits a kernel with~$\Oh(\ell^4)$ vertices. By using a polynomial-time~$2$-approximation algorithm for \feedbackvertexset~\cite{BeckerG96a}, we can again drop the assumption that such a set is supplied in the input.

\treewidthHierarchyPicture

After these two examples it becomes an interesting question whether there is a parameter even smaller than the feedback vertex number, in which the size of an instance can be bounded polynomially after an efficient preprocessing phase. The relationships between structural graph parameters given in \imgref{figure:treewidthHierarchyPicture} show that the deletion distance (i.e., the minimum size of a modulator) to a chordal graph is a smaller parameter than the minimum size of a feedback vertex set. Since \treewidth is trivially solvable on chordal graphs, the deletion distance to a chordal graph measures the ``distance from triviality''~\cite{Niedermeier10} of the instance. Accordingly one may hope that \treewidthByChordal is FPT and admits a polynomial kernel. Unfortunately, the existence of a polynomial kernel for this problem is very unlikely. We prove that even the larger parameterization by deletion distance to a single clique, fails to admit a polynomial kernel unless \containment. We use an intricate cross-composition~\cite{BodlaenderJK11} to establish this result, starting from the NP-completeness result for \treewidth by Arnborg et al.~\cite{ArnborgCP87}. Under the same assumption, our proof also shows that \pathwidthByClique does not admit a polynomial kernel.

When it comes to the weighted version of the problem, incompressibility sets in at a much higher level in the parameter hierarchy. We prove that \weightedtreewidth does not even admit a polynomial kernel when parameterized by the size of a vertex cover, unless \containment. Hence the presence of weights makes it difficult to preprocess instances of \treewidth, even when the graph has a very restricted structure. It is interesting to note the difference between \treewidth and \weightedtreewidth when  parameterized by vertex cover.

\paragraph{Organization of the paper.} After this introduction, we give preliminary definitions and lemmata in \sectref{section:preliminaries}. In \sectref{section:simplicial} we show that \treewidthByVC has a kernel with~$\Oh(\ell^3)$ vertices. To do so, we introduce a number of `safe' reduction rules: these are variants of rules from existing treewidth algorithms and preprocessing methods, including rules that remove simplicial vertices. In \sectref{section:fvs} we turn to \treewidthByFVS and show a kernel with~$\Oh(\ell^4)$ vertices. In addition to variants of the rules for vertex cover, a key role will be played by \emph{almost simplicial vertices}. We give a set of safe reduction rules that remove all such vertices. 

In \sectref{section:lowerbounds} we present our lower bound results. Our main lower bound, \thmref{theorem:twmsc:nokernel}, shows that \treewidthByClique does not admit kernels of polynomial size unless \containment. The theorem originates from our extended abstract on preprocessing for the \pathwidth problem~\cite{BodlaenderJK12a}. It strengthens the claims from a preliminary version of this work~\cite{BodlaenderJK11b}, where we only managed to prove kernel lower bounds for the parameterizations by deletion distance to a disjoint union of cliques (a cluster graph), and to the edge-complement of a cluster graph. As a single clique is both a cluster and a co-cluster graph, our present lower bound unifies both results. The lower bound for \weightedTreewidthByVC is given in \sectref{sect:tw:weightedLowerBound}. For completeness we establish in \sectref{sect:tw:fptClassifications} that the problems that we prove kernel lower bounds against are fixed-parameter tractable, and therefore admit kernels of exponential size. Some final remarks are made in \sectref{section:conclusions}.

\section{Preliminaries} \label{section:preliminaries} \label{section:definitions}
For a natural number~$n$ we use~$[n]$ as an abbreviation for the set~$\{1, 2, \ldots, n\}$. For a set~$X$ and natural number~$i$ the collection of all size-$i$ subsets of~$X$ is denoted~$\binom{X}{i}$. When analyzing the running time of an algorithm on a graph~$G$, we write~$n$ for the number of vertices and~$m$ for the number of edges.

\subsection{Graphs and separators}
In this work all graphs are finite, simple, and undirected. An undirected graph~$G$ consists of a vertex set~$V(G)$ and a set of edges~$E(G)$, whose elements are pairs of vertices. The open neighborhood of a vertex~$v\in V(G)$ in a graph~$G$ is denoted by~$N_G(v)$, and its closed neighborhood is~$N_G[v]$. If~$S \subseteq V(G)$ is a vertex set then~$G - S$ denotes the graph obtained from~$G$ by deleting all vertices of~$S$ and their incident edges. A vertex~$v$ is \emph{simplicial} in a graph~$G$ if~$N_G(v)$ is a clique. A vertex~$v$ is \emph{almost simplicial} in a graph~$G$ if~$v$ has a neighbor~$w$ such that~$N_G(v)-\{w\}$ is a clique. Such a vertex~$w$ is a \emph{special neighbor} of~$v$.

A set of vertices~$Q$ \emph{separates} vertices~$v$ and~$w$ if each path from~$v$ to~$w$ uses at least one vertex in~$Q$. A set of vertices is a \emph{separator} if there exist vertices~$v$ and~$w$ such that~$Q$ separates~$v$ from~$w$. The set~$Q$ \emph{minimally separates}~$v$ and~$w$ if it separates~$v$ and~$w$ but there is no proper subset of~$Q$ that separates~$v$ and~$w$. It is a \emph{minimal separator} if there is a pair of vertices~$v$ and~$w$ that are minimally separated by~$Q$. A set of vertices~$Q$ is a \emph{minimal almost clique separator}, if it is a minimal separator and there is a vertex~$v\in Q$ such that~$Q-\{v\}$ is a clique. By \emph{completing a vertex set~$S \subseteq V(G)$ of graph~$G$ into a clique} we mean the operation that adds edges between all pairs of~$S$ that were previously nonadjacent. By~$G_1\otimes G_2$ we denote the join of two graphs~$G_1$ and~$G_2$, obtained by taking their disjoint union and adding all edges between the two graphs. An edge contraction in a graph~$G$ is the operation where two adjacent vertices~$v$,~$w$ are replaced by one vertex with neighborhood~$(N_G(v) \cup N_G(w)) \setminus \{v,w\}$. A graph~$H$ is a \emph{minor} of a graph~$G$ if~$H$ can be obtained from~$G$ by a series of zero or more vertex deletions, edge deletions, and edge contractions. 

Recall that a graph is \emph{chordal} if all its cycles of length at least four have a chord, i.e., an edge between two vertices that are not successive on the cycle. A graph is \emph{bipartite} if its vertex set can be partitioned into two independent sets. A \emph{co-bipartite} graph is the edge-complement of a bipartite graph, and therefore admits a partition of its vertices into two cliques. For graph-theoretic notation not defined here we refer to one of the textbooks on the subject~\cite{BrandstadtLS99,Diestel10}.

\subsection{Treewidth} \label{section:treewidthdef}
A \emph{tree decomposition} of a graph~$G$ is a pair~$(T, \{\X_i \mid i\in V(T)\})$ with~$T$ a tree and~$\{\X_i \mid i\in V(T)\}$ a family of subsets of~$V(G)$, such that:
\begin{itemize}
\item $\bigcup_{i\in V(T)} \X_i = V(G)$.
\item For all~$\{v,w\}\in E(G)$, there is an~$i\in V(T)$ with~$v,w\in \X_i$.
\item For all~$v\in V(G)$, the set~$I_v = \{i\in V(T) \mid v\in \X_i\}$ induces a subtree of~$T$.
\end{itemize}
The sets~$\X_i$ are called the \emph{bags} of the tree decomposition. The \emph{width} of a tree decomposition~$(T, \{\X_i \mid i\in V(T)\})$ is~$\max_{i\in V(T)} |\X_i|-1$, and the \emph{treewidth} of~$G$ is the minimum width of a tree decomposition of~$G$.

Suppose we have a graph~$G$ with a weight function~$w \colon V(G) \rightarrow \mathbb{N}$. The \emph{weighted width} of a tree decomposition~$(T, \{\X_i \mid i\in V(T)\})$ of~$G$ is~$\max_{i\in V(T)} \sum_{v\in \X_i} w(v)$ minus one, and the \emph{weighted treewidth} of~$G$ is the minimum weighted width of a tree decomposition of~$G$.\footnote{Observe that we deviate from the definition of weighted width employed by Van den Eijkhof et al.~\cite{EijkhofBK07}, who do not subtract one from the maximum. Our definition ensures that the treewidth of a graph equals its weighted treewidth when all vertices have weight one.}

There are several alternative characterizations of treewidth, which are sometimes convenient to use in proofs. Let~$\omega(G)$ denote the size of a maximum clique in~$G$. The treewidth of a graph~$G$ is the minimum, over all chordal supergraphs~$H$ of~$G$, of~$\omega(H) - 1$ (see \cite[Theorem 1]{Bodlaender98}). 

Treewidth can also be characterized in terms of \emph{elimination orderings}. Accordingly, \emph{eliminating} a vertex~$v$ in a graph~$G$ is the operation of removing~$v$ while completing its open neighborhood into a clique. An elimination ordering of an $n$-vertex graph~$G$ is simply a permutation~$\pi \colon V(G) \to [n]$ of its vertices. Given an elimination ordering~$\pi$ for~$G$, we obtain a series of graphs by consecutively eliminating~$\pi^{-1}(1), \ldots, \pi^{-1}(n)$. The \emph{fill edges of~$\pi$} are the edges~$F_{\pi}$ that are added to the graph during this process when open neighborhoods are completed into cliques. The chordal supergraph of~$G$ on edge set~$E(G) \cup F_{\pi}$ is the \emph{filled graph} with respect to~$\pi$. If~$F_{\pi}$ is empty, then the elimination ordering is \emph{perfect}. The treewidth of a graph equals the minimum, over all elimination orderings of~$G$, of the maximum degree of a vertex at the time it is eliminated~\cite[Theorem 36]{Bodlaender98}. We will use the weighted 
analogue of this characterization in our kernelization lower bound of \sectref{sect:tw:distanceClique}.

\subsection{Lemmata}
We utilize a number of well-known facts to simplify our argumentation.

\begin{proposition}[See e.g., {\cite[Lemma 16]{Bodlaender98}}]\label{proposition:treewidthminor}
Let~$H$ be a minor of~$G$. Then the treewidth of~$H$ is at most the treewidth of~$G$.
\end{proposition}

\begin{proposition}[\cite{Dirac61}] \label{proposition:nonAdjacentSimplicial}
Every chordal graph that is not complete contains two nonadjacent simplicial vertices.
\end{proposition}

\begin{lemma}[See e.g., \cite{OlesenM02}]\label{lemma:cliqueseparator}
Let~$S\subseteq V(G)$ form a clique in~$G$. Suppose~$V_1, \ldots, V_r$ are the vertex sets of the connected components of~$G - S$. Then the treewidth of~$G$ equals the maximum treewidth of~$G[S\cup V_i]$ for~$i\in [r]$.
\end{lemma}

\begin{lemma}[Folklore]\label{lemma:pathcontainment}
Let~$G$ be a graph, and let~$(v_1, v_2, \ldots, v_r)$ be a path in~$G$. Let~$(T, \{\X_i \mid i\in V(T)\})$ be a tree decomposition of~$G$. Suppose~$i_1, i_2, i_3\in V(T)$ and~$i_2$ is on the path in~$T$ from~$i_1$ to~$i_3$. Suppose~$v_1\in \X_{i_1}$ and~$v_r\in \X_{i_3}$. Then~$\{v_1, v_2, \ldots, v_r\} \cap \X_{i_2} \neq \emptyset$.
\end{lemma}

\begin{lemma}\label{lemma:cliquecontainment2}
Let~$G$ be a graph, and let~$W_1, \ldots, W_r$ be sets of vertices such that:
\begin{itemize}
\item For each~$i \in [r]$ the set~$W_i$ induces a connected subgraph of~$G$.
\item For all $i,j \in [r]$ it holds that $W_i \cap W_j \neq \emptyset$ or~$W_i$ contains a vertex that is adjacent to a vertex in~$W_j$.
\end{itemize}
Then every tree decomposition of~$G$ has a bag containing at least one vertex of each set~$W_i$ for~$i \in [r]$.
\end{lemma}
\begin{proof}
Since each~$W_i$ for~$i \in [r]$ induces a connected subgraph, the bags containing a vertex in~$W_i$ form a subtree of~$T$ by \lemmaref{lemma:pathcontainment}. For each~$i$ and~$j$ in~$[r]$, there is a bag containing a vertex of~$W_i$ and a vertex of~$W_j$ by the second condition. Using these observations the lemma follows from the Helly property for subtrees of a tree~\cite[Definition 1.3.4]{BrandstadtLS99}.
\end{proof}

Vertex sets~$W_1, \ldots, W_r$ as described in \lemmaref{lemma:cliquecontainment2} form a \emph{bramble}. Brambles may be used to characterize treewidth. As we shall not use brambles further in this work, \lemmaref{lemma:cliquecontainment2} is stated in elementary terms. It easily implies the following well-known facts.

\begin{proposition} \label{proposition:simpleCliqueContainment}
If~$S$ is a clique in graph~$G$ then any tree decomposition for~$G$ has a bag containing all vertices of~$S$.
\end{proposition}

\begin{proposition} \label{proposition:treewidthOfClique}
The treewidth of a complete graph on~$n$ vertices is~$n-1$.
\end{proposition}

Finally, we use a result of Bodlaender and Koster~\cite{BodlaenderK06c} which shows that for the purpose of finding a minimum width tree decomposition one may safely complete any minimal almost clique separator into a clique.

\begin{theorem}[{\cite[Theorem 13 and Corollary 20]{BodlaenderK06c}}] \label{theorem:minimalalmostcliqueseparators}
If~$S$ is a minimal almost clique separator in~$G$, then~$\tw(G)=\tw(G')$ where~$G'$ is obtained by completing~$S$ into a clique. The set of all minimal almost clique separators of a graph can be found in~$\Oh(n^2 m)$ time.
\end{theorem}

%\section{Kernelization with respect to vertex cover number: eliminating simplicial vertices}
\section{A polynomial kernel for Treewidth parameterized by Vertex Cover}
\label{section:simplicial} \label{sect:tw:byVC}
The kernelization algorithms we present in this paper consist of a number of reduction rules. In each case, the input to the rule is a graph~$G$, an integer~$k$, and a deletion set~$S \subseteq V(G)$ such that~$G - S$ is a member of the relevant graph class \F, and the output is an instance~$(G', k', S')$. A rule is said to be \emph{safe} if for all inputs~$(G, k, S)$ which satisfy~$G - S \in \F$ we have~$\tw(G) \leq k \Leftrightarrow \tw(G') \leq k'$ and~$G' - S' \in \F$. We will sometimes say that the algorithm answers \yes or \no; this should be interpreted as outputting a constant-size \yes or \no instance of the problem at hand, i.e., a clique on three vertices with~$k=2$, respectively the same clique with~$k=1$.

In this section we show our kernelization for \treewidthByVC (i.e., parameterized by a modulator to an independent set). The kernelization focuses mostly on simplicial vertices; removing them (and possibly updating a bound for the treewidth) is a well-known and often used preprocessing rule for \treewidth; see the discussion of Bodlaender et al.~\cite{BodlaenderKE05}. Another rule, first used in the linear-time algorithm for computing treewidth~\cite{Bodlaender96} adds edges between vertices with many common neighbors. The rule was also used in lower bound heuristics for treewidth~\cite{Bodlaender2003,BodlaenderKW05,ClautiauxCMN03,ClautiauxMNC04}.

\begin{rul}[Low degree simplicial vertex]\label{rule:lowsimplicial}
If~$v$ is a simplicial vertex of degree at most~$k$ then remove~$v$.
\end{rul}

\begin{rul}[High degree simplicial vertex]\label{rule:highsimplicial}
If~$v$ is a simplicial vertex of degree greater than~$k$ then answer \no.
\end{rul}

Standard theory on treewidth shows that \rulesref{rule:lowsimplicial} and \ref{rule:highsimplicial} are safe. By \proposref{proposition:simpleCliqueContainment} the vertices of any clique in a graph~$G$ must occur together in at least one bag in any tree decomposition of~$G$. If the clique given by the closed neighborhood of a simplicial vertex is too big (\ruleref{rule:highsimplicial}) then we may reject; if it is sufficiently small, it is safe to add the vertex~$v$ later, since its neighbors occur in a shared bag.

It is well known that if non-adjacent vertices~$v,w$ have at least~$k+1$ common neighbors, then adding the edge~$\{v,w\}$ does not affect whether the treewidth of the graph is at most~$k$~\cite{Bodlaender96}: in any triangulation, either~$\{v,w\}$ is an edge, or all common neighbors of~$v$ and~$w$ form a clique. In the latter case, this clique plus~$v$ is a clique with at least~$k+2$ vertices, implying a treewidth of at least~$k+1$. We use this rule in a restricted setting, to ensure that~$S$ remains a vertex cover of the graph.

\begin{rul}[Common neighbors improvement]\label{rule:neighborimprovementS}
Suppose that~$\{v,w\}\not\in E(G)$ and that~$v\in S$ or~$w\in S$. If~$v$ and~$w$ have at least~$k+1$ common neighbors, then add the edge~$\{v,w\}$.
\end{rul}

Yet another simple rule is the following, using that~$S$ is a vertex cover of~$G$.

\begin{rul}[Trivial decision]\label{rule:trivial1}
If~$k\geq |S|$, then answer \yes.
\end{rul}

Safeness can be argued as follows. The treewidth of~$G$ is at most~$|S|$: for each~$v \in V(G) \setminus S$, take a bag with vertex set~$S\cup\{v\}$, and connect these bags in any way. This gives a tree decomposition of~$G$ of width at most~$|S|$.

It is not hard to argue the, possibly surprising, fact that the exhaustive application of \rulesref{rule:lowsimplicial}--\ref{rule:trivial1} (i.e., until we answer \no or \yes, or no application of one of these rules is possible) already gives a polynomial kernel for \treewidthByVC. It is clear that this reduction can be performed in polynomial time; an algorithm with runtime~$\Oh(nm)$ is easy to obtain.

\begin{theorem}\label{theorem:vckernel}
\treewidthByVC has a kernel with~$\Oh(\ell^3)$ vertices.
\end{theorem}

\begin{proof}
Let~$(G,k,S)$ be an instance of \treewidthByVC. Let~$(G',k',S')$ be the instance obtained from exhaustive application of \rulesref{rule:lowsimplicial}--\ref{rule:trivial1}. By safety of the reduction rules~$(G',k',S')$ is \yes if and only if~$(G,k,S)$ is \yes.

The reduction rules guarantee that~$S' \subseteq S$ is a vertex cover in~$G'$, with~$|S'| \leq |S| = \ell$. Each vertex~$v \in V(G') \setminus S'$ has at least one pair of distinct neighbors in~$S'$ that are not adjacent, otherwise~$v$ is simplicial and would have been handled by \ruleref{rule:lowsimplicial} or \ruleref{rule:highsimplicial}. Assign~$v$ to this pair. If we assign~$v$ to the pair~$\{w,x\}$, then~$v$ is a common neighbor of~$w$ and~$x$. Hence a pair cannot have more than~$k$ vertices assigned to it, otherwise \ruleref{rule:neighborimprovementS} applies. As there are at most~$\binom{\ell}{2}$ pairs of non-adjacent neighbors in~$S'$, we have~$|V(G') \setminus S'| \leq k \cdot \binom{\ell}{2} \leq \ell \cdot \binom{\ell}{2} \in \Oh(\ell^3)$. The last step uses the fact that~$k \leq \ell$ by \ruleref{rule:trivial1}.
\end{proof}

By combining \thmref{theorem:vckernel} with a polynomial-time $2$-approximation algorithm for vertex cover, we obtain the following corollary.
\begin{corollary}
There is a polynomial-time algorithm that given an instance $(G,k)$ of \treewidth computes an equivalent instance $(G',k)$ such that~$V(G') \subseteq V(G)$ and~$|V(G')| \in \Oh(\vc(G)^3)$, where~$\vc(G)$ is the size of a minimum vertex cover of~$G$.
\end{corollary}

\section{A polynomial kernel for Treewidth parameterized by Feedback Vertex Set} \label{section:fvs} \label{sect:tw:byFVS}
In this section, we establish that \treewidthByFVS (of size~$\ell$) has a kernel with~$\Oh(\ell^4)$ vertices. The kernelization algorithm is again given by a set of safe reduction rules, that are applied while possible: three simple rules, three rules that remove all almost simplicial vertices (\sectref{subsection:almostsimplicial}), two rules that reduce the graph when there is a clique-seeing path (defined in \sectref{section:cliqueseeingpaths:rules}), and a cut-off rule that rejects when there are long clique-seeing paths left after exhaustive application of the earlier rules (\sectref{subsection:cutoffrule}). In \sectref{section:counting}, we show that graphs to which no rule applies have $\Oh(\ell^4)$ vertices, and thus arrive at our kernel bound.

The first new rule generalizes \ruleref{rule:neighborimprovementS}; it was used in experiments by Clautiaux et al.~\cite{ClautiauxCMN03} and its correctness is proven by Bodlaender~\cite[Lemma 5]{Bodlaender2003}. The rule can be implemented in polynomial time by using a maximum flow algorithm to find the disjoint paths.

\begin{rul}[Disjoint paths improvement]\label{rule:neighborimprovementR}
Suppose~$\{v,w\}\not\in E(G)$ and that~$v\in S$ or~$w\in S$. If there are at least~$k+1$ internally vertex-disjoint paths between~$v$ and~$w$, then add the edge~$\{v,w\}$.
\end{rul}

The next rule also adds edges that do not affect the answer to the problem. It is based upon the notion of \emph{minimal almost clique separators} of Bodlaender and Koster \cite{BodlaenderK06c} (see \sectref{section:preliminaries}). They show that the treewidth is not changed when edges are added to complete a minimal almost clique separator into a clique. We use a version of this rule which ensures that~$S$ continues to be a feedback vertex set. Safeness of the following rule and an efficient implementation follow directly from \thmref{theorem:minimalalmostcliqueseparators}.

\begin{rul}\label{rule:almostcliqueseparator}
Let~$Q$ be a minimal almost clique separator in~$G$, and suppose there is at most one vertex~$v\in Q$ with~$v\not\in S$. Then complete~$Q$ into a clique by adding edges between each pair of non-adjacent vertices in~$Q$.
\end{rul}

The third new rule is straight forward, and is correct because reasoning similar to that of \ruleref{rule:trivial1} shows that each graph with a feedback vertex set of size~$\ell$ has treewidth at most~$\ell+1$~\cite[Corollary 75]{Bodlaender98}.
\begin{rul}[Trivial decision]\label{rule:trivialfvs}
If~$k \geq |S| + 1$, then answer \yes.
\end{rul}

\subsection{Almost simplicial vertices} \label{subsection:almostsimplicial}
Bodlaender et al.~\cite{BodlaenderKE05} introduced the notion of an \emph{almost simplicial vertex}, and gave a reduction rule that removes almost simplicial vertices whose degree is at most a known lower bound for the treewidth of the input graph. In this section we give a set of rules that can also remove almost simplicial vertices of higher degree. \ruleref{rule:lowalmostsimplicial} is a reformulation of the existing \emph{Low Degree Almost Simplicial Vertex Rule} by Bodlaender et al.~\cite{BodlaenderKE05}. \ruleref{rule:highalmostsimplicial} gives a simple way to deal with almost simplicial vertices of degree larger than~$k+1$. Our novel reduction rule is given as \ruleref{rule:kp1almostsimplicial}; it deals with almost simplicial vertices of degree exactly~$k+1$.

\begin{rul}[Low Degree Almost Simplicial Vertex]\label{rule:lowalmostsimplicial}
Let~$v$ be an almost simplicial vertex with special neighbor~$w$. If the degree of~$v$ is at most~$k$, then contract the edge~$\{v,w\}$ into~$w$ obtaining~$G'$. If~$v\in S$, then let~$S':=S\setminus\{v\}\cup\{w\}$, else let~$S':=S$.
\end{rul}
\begin{lemma}
\ruleref{rule:lowalmostsimplicial} is safe.
\end{lemma}
\begin{proof}
It is clear that~$S'$ is a feedback vertex set of~$G'$. Let~$G'$ be the graph resulting after the operation. If the treewidth of~$G$ is at most~$k$, then the treewidth of~$G'$ is at most~$k$ as the treewidth cannot increase by contraction (\proposref{proposition:treewidthminor}).

Suppose the treewidth of~$G'$ is at most~$k$. Take a tree decomposition of~$G'$ of width at most~$k$. As~$N_G(v)$ is a clique in~$G'$, by \proposref{proposition:simpleCliqueContainment} there is a bag that contains all vertices of~$N_G(v)$, say~$N_G(v) \subseteq \X_i$. Add a new bag with vertex set~$N_G[v]$ and make it adjacent in the tree decomposition to node~$i$; we obtain a tree decomposition of~$G$ of width at most~$k$.
\end{proof}

\begin{rul}[High Degree Almost Simplicial Vertex]\label{rule:highalmostsimplicial}
Let~$v$ be an almost simplicial vertex. If the degree of~$v$ is at least~$k+2$,
then answer \no.
\end{rul}

Correctness is obvious:~$v$ with its neighbors except its special neighbor forms a clique with at least~$k+2$ vertices, so the treewidth is larger than~$k$. We introduce a new, more complex rule that deals with almost simplicial vertices of degree exactly~$k+1$. 

\begin{rul}[Degree~$k+1$ Almost Simplicial Vertex]\label{rule:kp1almostsimplicial}
Let~$v$ be an almost simplicial vertex with special neighbor~$w$, and let
the degree of~$v$ be exactly~$k+1$.
\begin{itemize}
\item If for each vertex~$x\in N_G(v) \setminus \{w\}$, there is an edge~$\{x,w\}\in E(G)$ or a path in~$G$ from~$x$ to~$w$ that avoids~$N_G[v] \setminus \{x,w\}$, answer \no.
\item Otherwise, contract the edge~$\{v,w\}$ to a new vertex~$x$, obtaining~$G'$. If~$v\in S$ or~$w\in S$, then let~$S':=S\setminus\{v,w\}\cup\{x\}$, else let~$S':=S$.
\end{itemize}
\end{rul}

\begin{figure}[t]
\centering
\includegraphics[scale=0.9]{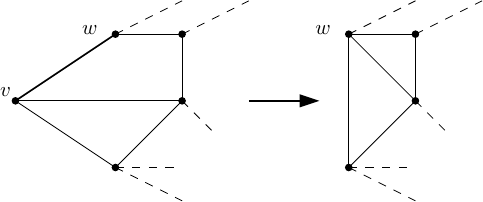}
\caption{\label{figure:degreek1almostsimplicial}An example of an application of the \emph{Degree~$k+1$ Almost Simplicial Vertex Rule}, second case ($k = 2$).}
\end{figure}

An example of an application of the second case of the \emph{Degree~$k+1$ Almost Simplicial Vertex Rule} is given in \imgref{figure:degreek1almostsimplicial}. It is easy to see that~$S'$ is a feedback vertex set for~$G'$. To argue safeness of \ruleref{rule:kp1almostsimplicial} we need two intermediate lemmata. The first one shows that deciding \no in the first case is indeed correct.

\begin{lemma}\label{lemma:safek+1-1}
Let~$v$ be an almost simplicial vertex with special neighbor~$w$, and let the degree of~$v$ be exactly~$k+1$. Suppose that for each vertex~$x\in N_G(v) \setminus \{w\}$, there is an edge~$\{x,w\}\in E(G)$ or a path in~$G$ from~$x$ to~$w$ that does not use any vertex in~$N_G[v] \setminus \{x,w\}$. Then the treewidth of~$G$ is at least~$k+1$.
\end{lemma}

\begin{proof}
We build a clique minor of~$G$ with~$k+2$ vertices. Consider the connected components of~$G-N_G[v]$. Each connected component that has no vertex adjacent to~$w$ is removed. Each connected component with a vertex adjacent to~$w$ is contracted to~$w$. We claim that this gives a clique on the vertex set~$N_G[v]$, i.e.,~$G$ has a clique with~$k+2$ vertices as a minor, and hence has treewidth at least~$k+1$ by \proposref{proposition:treewidthminor} and  \proposref{proposition:treewidthOfClique}.

Note that the only edges that are missing in~$G[N_G[v]]$ are edges between vertices~$x\in N_G(v) \setminus \{w\}$ and~$w$. If~$\{x,w\}\not\in E(G)$, then there is a path from~$x$ to~$w$ in~$G$ that avoids~$N_G[v] \setminus \{x,w\}$. Hence, this path must belong to a connected component of~$G-N_G[v]$ that is connected to~$w$,
and thus, the edge~$\{x,w\}$ is present in the constructed minor. This shows the claim, and thus the lemma.
\end{proof}

For showing the safeness of the second step we use \lemmaref{lemma:cliqueseparator}.

\begin{lemma}\label{lemma:safek+1-2}
Let~$v$ be an almost simplicial vertex with special neighbor~$w$. Suppose that there is a vertex~$z\in N_G(v)$ with~$z\neq w$ such that~$N_G[v] \setminus \{z,w\}$ separates~$z$ from~$w$. Then the treewidth of~$G$ is at most~$k$, if and only if the graph~$G'$ obtained by contracting the edge~$\{v,w\}$ into~$w$ has treewidth at most~$k$.
\end{lemma}
\begin{proof}
If the treewidth of~$G$ is at most~$k$ then the treewidth of its minor~$G'$ is at most~$k$ (\proposref{proposition:treewidthminor}). It suffices to show the converse. Assume therefore that~$G'$ has treewidth at most~$k$.

Consider the set~$Z := N_G[v] \setminus \{z,w\}$. By the preconditions to the lemma~$Z$ is a clique in~$G$ that separates~$z$ from~$w$. Let~$V_1, \ldots, V_r$ be the connected components of~$G - Z$. Assume without loss of generality that~$V_1$ contains the vertex~$w$. By the separation property we know that~$z \not \in V_1$.

Applying \lemmaref{lemma:cliqueseparator} to the clique separator~$Z$ with respect to graph~$G$ we find that~$\tw(G) = \max_{i \in [r]} \tw(G[V_i \cup Z])$. For~$i \in \{2, \ldots, r\}$ it is easy to see that~$G[V_i \cup Z]$ is a subgraph of~$G'$ and therefore has treewidth at most~$k$. To establish that~$G$ has treewidth at most~$k$, it remains to show that~$G_1 := G[V_1 \cup Z]$ has treewidth at most~$k$.

Consider the graph~$G'_1 := G'[V_1 \cup (Z \setminus \{v\}) \cup \{z\}]$. It contains a vertex labeled~$w$ (the result of the contraction), and it contains vertex~$z$, but it does not have vertex~$v$. As~$G'_1$ is a subgraph of~$G'$, it has treewidth at most~$k$. We will show that~$G'_1$ is a supergraph of~$G_1$, resulting in the desired treewidth bound for~$G_1$.

Observe that~$N_{G'_1}(z) \supseteq N_{G_1}(v)$: in graph~$G$, the vertex~$z$ is a neighbor of~$v$ different from the special neighbor; hence it is adjacent to all vertices in~$N_G(v) \setminus \{z,w\}$, and this remains true when transforming into~$G'$. Since~$\{v,z\} \in E(G)$ the contraction of edge~$\{v,w\}$ ensures that in~$G'$ (and therefore~$G'_1$), vertex~$z$ is also adjacent to~$w$. Thus it indeed holds that~$N_{G'_1}(z) \supseteq N_{G_1}(v) = N_G(v) \setminus \{z\}$. As~$z \not \in V(G_1)$ one may now verify that relabeling vertex~$z$ to~$v$ in graph~$G'_1$ yields a supergraph of~$G_1$. This shows that~$\tw(G_1) \leq \tw(G'_1) \leq k$ and concludes the proof.
\end{proof}

Safeness of \ruleref{rule:kp1almostsimplicial} now follows from \lemmataref{lemma:safek+1-1} and \ref{lemma:safek+1-2}. Note that the rules for almost simplicial vertices (\rulesref{rule:lowalmostsimplicial},~\ref{rule:highalmostsimplicial}, and~\ref{rule:kp1almostsimplicial}) can be easily seen to subsume the rules for simplicial vertices (\rulesref{rule:lowsimplicial} and~\ref{rule:highsimplicial}) used in the previous section. In particular, the first case of \ruleref{rule:kp1almostsimplicial} covers simplicial vertices of degree~$k+1$.

We point out that \rulesref{rule:lowalmostsimplicial},~\ref{rule:highalmostsimplicial}, and~\ref{rule:kp1almostsimplicial} can be efficiently implemented since they come down to checking whether a vertex is almost simplicial plus checking reachability (in \ruleref{rule:kp1almostsimplicial}).
The following proposition gives a bound on the number of leaves in~$G-S$ in reduced instances.

% The following proposition follows from counting arguments similar to those in the proof of \thmref{theorem:vckernel}, by observing that after exhaustive application of the rules no leaf in the forest~$G - S$ is almost simplicial, and hence every such leaf must have a pair of neighbors in~$S$ that are non-adjacent.

\begin{proposition}\label{proposition:tja}
Suppose an instance~$(G,k)$ of \treewidth is given together with a feedback vertex set~$S$ of size~$\ell$. If we exhaustively apply \rulesref{rule:neighborimprovementR}--\ref{rule:kp1almostsimplicial}, then we obtain in polynomial time an equivalent instance~$(G', k)$ with a feedback vertex set~$S'$ of size at most~$\ell$, such that the forest~$G'-S'$ has~$\Oh(\ell^3)$ leaves.
\end{proposition}

\begin{proof}
Consider any leaf~$v$ in~$G'-S'$. Since no reduction rule applies to~$(G',k)$ with feedback vertex set~$S'$ it follows that~$v$ cannot be simplicial or almost simplicial. As~$v$ has (at most) one neighbor in~$G'-S'$ it would be almost simplicial if its remaining neighbors, which are all in~$S'$, form a clique. In other words, since~$(G',k)$ is exhaustively reduced, for every leaf~$v$ of~$G'-S'$ there must be two nonadjacent neighbors~$u$ and~$w$ in~$S'$. Since there are less than~$|S'|^2\in\Oh(\ell^2)$ pairs~$\{u,w\}\subseteq S'$ and each pair has at most~$k$ such shared neighbors~$v$ (by \ruleref{rule:neighborimprovementR}) this gives a bound of~$\Oh(\ell^3)$ on the number of leaves of~$G'-S'$; recall that~$k\leq \ell$ by \ruleref{rule:trivialfvs}.
\end{proof}

\paragraph{Preprocessing heuristics. }
At this point we digress to discuss the relationship of the rules presented in this section with earlier work. Bodlaender et al.~\cite{BodlaenderKE05} gave an \emph{Almost Simplicial Vertex Rule} as a \emph{preprocessing heuristic} for the optimization version of \treewidth. They used a variable \low as a lower bound on the treewidth of the original input graph; the rule could be applied to almost simplicial vertices of degree at most \low. Thus, the earlier preprocessing rules could not deal with almost simplicial vertices of degree at least $\low+1$. Using a variant of \ruleref{rule:kp1almostsimplicial}, we can now extend the \emph{Almost Simplicial Vertex Rule} and preprocess a graph such that we remove all vertices that are almost simplicial, as follows. Suppose~$v$ is an almost simplicial vertex of degree~$d$ with special neighbor~$w$. If~$d-1 > \low$, then \low is set to~$d-1$. Then, if~$d-1 = \low$, we check if for each vertex~$x\in N_G(v)-\{w\}$, there is an edge~$\{x,w\}\in E(G)$ or a path from~$x$ to~$w$ that does not use any vertex in~$N_G[v]-\{x,w\}$. If so, \low is increased by one. Then we contract the edge~$\{x,w\}$. This rule is \emph{safe} in the sense that the treewidth of the original graph~$G$ equals the maximum of \low and the treewidth of the reduced graph.

\subsection{Clique-seeing paths}\label{section:cliqueseeingpaths}
The reduction rules presented so far ensure that the number of leaves in the forest~$G-S$ is cubic in the size of the modulator~$S$. Since the size of a forest can be bounded polynomially in the number of leaves and the maximum number of consecutive vertices of degree two, we can obtain a polynomial kernel by bounding the length of degree-two paths in the forest.

\ruleref{rule:neighborimprovementR} ensures that for each pair of non-adjacent vertices~$u$ and~$v$ in~$S$, there are at most~$k$ vertices in the forest that see both~$u$ and~$v$. Hence the number of vertices that do not see a clique in~$S$ is~$\Oh(k \cdot \ell^2) \leq \Oh(\ell^3)$, and it suffices to bound the length of paths in the forest~$G-S$ that consist only of vertices seeing a clique in~$S$. The following notion will therefore be the key to the remaining reduction steps.

\begin{definition}
We call a path~$(v_0, v_1, \ldots, v_r, v_{r+1})$ in~$G$ a \emph{clique-seeing path} that \emph{sees} clique~$X$ if the following hold.
\begin{itemize}
\item $X = \bigcup_{i\in[r]} N_G(v_i) \setminus \{v_0, v_1, \ldots, v_{r+1}\}$ is a clique.
\item For each~$i \in [r]$ it holds that $N_G(v_i) \subseteq \{v_{i-1},v_{i+1}\} \cup X$.
\end{itemize}
\end{definition}

\begin{figure}[t]
\centering
\includegraphics[scale=1.0]{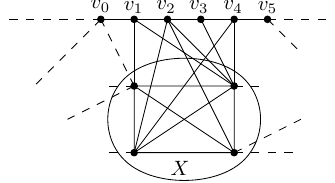}
\caption{\label{figure:cliqueseeingpath}An example of a clique-seeing path.}
\end{figure}

\noindent An example is given in \imgref{figure:cliqueseeingpath}. Note that~$v_0$ and~$v_{r+1}$ play a special role. Each vertex~$v_i$ with~$i \in [r]$ has
exactly two neighbors outside~$X$, namely the previous and next vertex on the path ($v_{i-1}$ and~$v_{i+1}$). Observe that even a path~$(v_0, v_1, \ldots, v_r, v_{r+1})$ in which~$N_G(v_i) \setminus \{v_{i-1}, v_{i+1}\}$ is a clique for each~$i \in [r]$, might not be clique-seeing. For example, if these sets are disjoint for all~$i \in [r]$ then even though the individual sets are cliques, the sets for different choices of~$i$ might not be adjacent to each other; but the definition of a clique-seeing path requires that the \emph{union} of the vertices seen by the path forms a clique. Paths that fail to be clique-seeing for this reason are not a big obstacle in the kernelization: as each such path provides a connection between a pair of nonadjacent vertices, \ruleref{rule:neighborimprovementR} is triggered if there are many such paths.

Our remaining task for the kernelization is to reduce the lengths of clique-seeing paths, without changing the answer to the problem. In some cases, however, the presence of a long clique-seeing path causes the treewidth of the graph to exceed~$k$. For example, this happens when the path, together with the clique that it sees, forms a $k+2$-clique minor. Our transformation rules will not shrink such paths; rather, we will detect their presence and outright answer~\no. The following lemma therefore sets up sufficient conditions for finding a clique minor of size~$k+2$. Although we will not need the lemma until \sectref{subsection:cutoffrule}, where we introduce the cutoff rule that answers \no, we present it here to give some intuition of the importance of clique-seeing paths.

\begin{lemma}\label{lemma:largeX}
Let~$(v_0, \ldots, v_{r+1})$ be a clique-seeing path, seeing clique~$X$. Suppose that~$v_0$ and~$v_{r+1}$ belong to the same connected component of~$G-(\{v_1,\ldots, v_r\} \cup X)$.
Let~$Z$ be the vertex set of the connected component of~$G- (\{v_1,\ldots,v_r\} \cup X)$ that contains~$v_0$. If one of the following cases holds, then~$G$ has treewidth at least~$k+1$.
\begin{enumerate}
\item $|X|\geq k+1$.
\item $|X|=k$ and each vertex in~$X$ has a neighbor in~$Z$.
\item $|X|=k-1$ and each vertex in~$X$ has a
neighbor in~$Z$, and there is an~$\ell \in [r-1]$, such that each vertex in~$X$ is adjacent to at least one vertex in~$\{v_1, \ldots, v_\ell\}$, and each vertex in~$X$ is adjacent to at least one vertex in~$\{v_{\ell+1}, \ldots, v_r\}$.
\end{enumerate}
\end{lemma}

\begin{proof}
In each case~$G$ contains a clique of size~$k+2$ as a minor, and hence has treewidth at least~$k+1$, by \proposref{proposition:treewidthminor}. In the first case, contract the interior~$\{v_1, \ldots, v_r\}$ of the path to a single vertex; it then forms a clique with the vertices in~$X$. In the second case, contract the interior of the path to a single vertex, and contract~$Z$ to a single vertex; these form a clique with the vertices in~$X$. In the last case, take~$X$, and contract each of the following to a single vertex: all vertices in~$Z$,~$\{v_1, \ldots, v_\ell\}$, and~$\{v_{\ell+1}, \ldots,v_r\}$.
\end{proof}

The lemma shows that when the endpoints~$v_0$ and~$v_{r+1}$ of a clique-seeing path are not separated by the union of the internal vertices and the seen clique, the path between~$v_0$ and~$v_{r+1}$ that avoids this separator is useful for building a large clique minor. The next lemma considers the case that~$v_0$ and~$v_{r+1}$ \emph{are} separated by~$X \cup \{v_1, \ldots, v_r\}$. In that case, the exhaustive application of \ruleref{rule:almostcliqueseparator} has a useful effect on separators formed by the path and the seen clique.

\begin{lemma} \label{lemma:separatorsOnCliqueSeeingPath}
Let~$(v_0, v_1, \ldots, v_r, v_{r+1})$ be a clique-seeing path in~$G$ that sees clique~$X  \subseteq S$, such that~$X \cup \{v_1, \ldots, v_r\}$ separates~$v_0$ from~$v_{r+1}$. If \ruleref{rule:almostcliqueseparator} cannot be applied, then for each~$i \in [r]$ the set~$\{v_i\} \cup (N_G(v_i) \cap X)$ is a clique separator that separates~$v_{i-1}$ from~$v_{i+1}$.
\end{lemma}
\begin{proof}
Consider a clique-seeing path that satisfies the stated requirements, and let~$i \in [r]$ be arbitrary. As~$X$ is a clique, the set~$\{v_i\} \cup (N_G(v_i) \cap X)$ is a clique; it remains to prove that it separates~$v_{i-1}$ from~$v_{i+1}$ if the rule is not applicable.

From the structure of the clique-seeing path and the fact that~$X \cup \{v_1, \ldots, v_r\}$ separates~$v_0$ from~$v_{r+1}$, it easily follows that~$\{v_i\} \cup X$ separates~$v_{i-1}$ from~$v_{i+1}$. Since~$X$ is a clique, the set~$\{v_i\} \cup X$ is almost a clique. As we just established that it is a~$v_{i-1}-v_{i+1}$ separator, it follows that it is an almost clique separator. Therefore the set~$\{v_i\} \cup X$ contains a minimal almost clique separator~$Z$ that separates~$v_{i-1}$ from~$v_{i+1}$, and~$Z$ must contain~$v_i$ because it is a common neighbor of the separated vertices. If~$Z$ is a clique, then~$Z \subseteq \{v_i\} \cup (N_G(v_i) \cap X)$ which proves that the superset~$\{v_i\} \cup (N_G(v_i) \cap X)$ is also a~$v_{i-1}-v_{i+1}$ separator (and hence a clique separator) and we are done.

Now assume that~$Z$ is not a clique. Since the clique~$X$ is a subset of the feedback vertex set~$S$, we know that the minimal almost clique separator~$Z$ contains at most one vertex which is not from~$S$, and therefore \ruleref{rule:almostcliqueseparator} is applicable. This proves the claim.
\end{proof}

\lemmaref{lemma:separatorsOnCliqueSeeingPath} will be useful when proving the correctness of the rules we present, and when analyzing their effects. 

\subsection{Reduction rules for clique-seeing paths}\label{section:cliqueseeingpaths:rules}
Having studied some of the properties of clique-seeing paths, we now turn to the reduction rules that act on them. The first rule in this section deals with clique-seeing paths that see a clique of size at most~$k-2$. The second considers the case that~$\{v_1, \ldots, v_r\} \cup X$ separates~$v_0$ and~$v_{r+1}$ in the graph. As the correctness proofs for these rules are fairly detailed and technical, we postpone their proofs to the end of the section. In our analysis it is sufficient to consider clique-seeing paths in which all vertices on the path reside in the forest~$G-S$, and all vertices in the seen clique belong to the feedback vertex set~$S$, but the rules are also safe in other cases.

\begin{rul}\label{rule:subsetpath}
Suppose we have a clique-seeing path~$(v_0, v_1, \ldots, v_r, v_{r+1})$ in~$G$, whose vertices avoid~$S$, that sees a clique~$X \subseteq S$ with $|X|\leq k-2$. If
\begin{equation} \nonumber
N_G(v_r)\cap X \subseteq \bigcup_{i \in [r-1]} N_G(v_i)\cap X
\end{equation}
then contract the edge~$\{v_r, v_{r+1}\}$ into the vertex~$v_{r+1}$, obtaining~$G'$.
\end{rul}

The rule contracts vertex~$v_r$, on a clique-seeing path that sees a small clique, into its neighbor~$v_{r+1}$, when all clique-neighbors of~$v_r$ are also seen by an earlier vertex on the path. The requirement that~$|X| \leq k-2$ is crucial for the correctness of the rule. To see this, consider the setting where we have a clique~$X$ of size~$k-1$ and a path~$(v_0, v_1, v_2, v_3)$ such that~$v_0$ sees no vertex of~$X$, while~$\{v_1,v_2,v_3\}$ see all vertices of~$X$. In such situations, the existence of a path from~$v_0$ to~$v_3$ avoiding~$X \cup \{v_1, v_2\}$ implies that the treewidth is at least~$k+1$: contracting the vertices on this path into~$v_3$ results in a clique minor on the vertex set~$\{v_1, v_2, v_3\} \cup X$ of size~$|X| + 3 = k+2$. Applying the rule to this structure would not be safe, as contracting the edge~$\{v_r, v_{r+1}\}$ in this situation destroys the clique minor and can cause the treewidth to drop to~$k$.

This example gives another indication that for clique-seeing paths, it is important whether the set~$X \cup \{v_1, \ldots, v_r\}$ separates the extremal vertices on the path~$\{v_0, v_{r+1}\}$ from each other. The next rule shows that if this is the case, we can reduce clique-seeing paths even if they see a large clique. Exhaustive application of the rule will provide us with a useful connectivity property that allows a rejection of instances with clique-seeing paths that remain long after application of \rulesref{rule:neighborimprovementR} to~\ref{rule:cliqueseeingsep}.

\begin{rul}\label{rule:cliqueseeingsep}
Suppose~$(v_0,v_1, v_2, v_3, v_4)$ is a clique-seeing path in~$G$, whose vertices avoid~$S$, that sees clique~$X \subseteq S$. Suppose~$\{v_1,v_2,v_3\}\cup X$ separates~$v_0$ from~$v_4$ and suppose that \ruleref{rule:almostcliqueseparator} cannot be applied. Compute the treewidth of~$G[\{v_1,v_2,v_3\} \cup X]$. If it is larger than~$k$, then answer \no, otherwise remove~$v_2$ from~$G$.
\end{rul}

It is not immediately obvious that given a path~$(v_0, v_1, \ldots, v_r, v_{r+1})$ \ruleref{rule:cliqueseeingsep} can be applied in polynomial time, when applicable. Since it is easy to verify whether the neighbors seen by the path form a clique, and whether~$v_0$ and~$v_{r+1}$ are separated, the crucial part is to compute the treewidth of~$G[\{v_1,v_2,v_3\} \cup X]$. The next lemma shows how to do this.

\begin{lemma}
There is a polynomial-time algorithm for deciding whether~$G[\{v_1,v_2,v_3\} \cup X]$ has treewidth at most~$k$.
\end{lemma}

\begin{proof}
Let us first observe that~$v_1$ and~$v_3$ are almost simplicial in~$G[\{v_1,v_2,v_3\}\cup X]$. Thus, by two applications of the rules for almost simplicial vertices (\rulesref{rule:lowalmostsimplicial}--\ref{rule:kp1almostsimplicial}) we either get \no (since the treewidth exceeds~$k$) or a clique (with treewidth clique size minus one). In both cases we efficiently answer whether~$\tw(G[\{v_1,v_2,v_3\} \cup X])\leq k$.
\end{proof}

Thus \ruleref{rule:cliqueseeingsep} can be implemented in polynomial time. From exhaustive application of the rule, we obtain a connectivity condition on the endpoints of clique-seeing paths which is captured by the following proposition.

\begin{proposition}\label{proposition:connectedness}
If \ruleref{rule:almostcliqueseparator} and \ruleref{rule:cliqueseeingsep} cannot be applied, and there is a clique-seeing path $(v_0, v_1, \ldots, v_r, v_{r+1})$ with~$r \geq 3$, whose vertices avoid~$S$, which sees a clique~$X \subseteq S$, then~$v_0$ and~$v_{r+1}$ are connected via a path that avoids~$\{v_1, \ldots, v_r\} \cup X$.
\end{proposition}
\begin{proof}
Assume that~$(v_0, v_1, \ldots, v_r, v_{r+1})$ with~$r \geq 3$ avoids~$S$ and sees the clique~$X \subseteq S$, and that the set~$\{v_1, \ldots, v_r\} \cup X$ separates~$v_0$ from~$v_{r+1}$. Consider the clique-seeing subpath~$(v_0, v_1, v_2, v_3, v_4)$ that sees a (not necessarily strict) subset~$X'$ of the clique~$X$. By \lemmaref{lemma:separatorsOnCliqueSeeingPath} the set~$\{v_1\} \cup (N_G(v_1) \cap X)$ separates~$v_{0}$ from~$v_{2}$, and therefore separates~$v_0$ from~$v_{4}$. As~$X' \supseteq N_G(v_1) \cap X$ the set~$\{v_1, v_2, v_3\} \cup X'$ also separates~$v_0$ from~$v_{4}$; but then \ruleref{rule:cliqueseeingsep} is applicable, a contradiction.
\end{proof}

\subsubsection{Correctness of Rule \ref{rule:subsetpath}}

The intuition behind the correctness proof for \ruleref{rule:subsetpath} is that a vertex~$v_r$ fulfilling the requirements can always be added at no cost in bag size or treewidth, by adding an additional bag. The detailed argumentation is unfortunately rather long, since the rule requires only a weak restriction of the graph structure; accordingly, there are many ways in which a tree decomposition for the resulting graph can be structured. Basically, we need to argue about the existence of certain bags containing the neighbors of~$v_r$ and draw conclusions that lead us to a good place for modifying the decomposition and appending bag(s) to place~$v_r$.

\begin{lemma}\label{lemma:subsetpathrulesafe}
\ruleref{rule:subsetpath} is safe.
\end{lemma}

\begin{proof}
For convenience, let us first recall the setting of \ruleref{rule:subsetpath}. It assumes that we have a clique-seeing path~$(v_0, v_1, \ldots, v_r, v_{r+1})$ in~$G$, whose vertices avoid~$S$, that sees a clique~$X \subseteq S$. Then, if~$|X|\leq k-2$ and
\begin{equation} \nonumber
N_G(v_r)\cap X \subseteq \bigcup_{i \in [r-1]} N_G(v_i)\cap X,
\end{equation}
we contract the edge~$\{v_r, v_{r+1}\}$ into the vertex~$v_{r+1}$, obtaining~$G'$.

As the path avoids~$S$, the contraction occurs in the forest~$G-S$ and therefore the resulting graph~$G' - S$ is still a forest; consequently, the set~$S$ is a feedback vertex set in~$G'$. Let~$v_{r+1}$ be the name of the vertex resulting from the contraction. As~$G'$ is a minor of~$G$, \proposref{proposition:treewidthminor} shows that~$\tw(G) \leq k$ implies that~$\tw(G') \leq k$.

Now suppose that the treewidth of~$G'$ is at most~$k$. The remainder of this proof will show that the treewidth of~$G$ is at most~$k$. Consider a tree decomposition~$(T, \{\X_i \mid i\in V(T)\})$ of~$G'$ of width at most~$k$. Recall from the definition of a clique-seeing path that~$X$, the vertices that do not lie on path~$(v_0, v_1, \ldots, v_r, v_{r+1})$ but are adjacent to an internal vertex of that path, form a clique. % Write~$Y = \bigcup_{i \in [r-1]} N_G(v_i)\cap X$. 
We show that the decomposition contains three special bags that will be useful to obtain a tree decomposition of~$G$.

Using \lemmaref{lemma:cliquecontainment2} (with one set for~$\{v_1, \ldots, v_{r-1}\}$, and one set for each vertex in~$X$) it follows that there is a bag~$i_1\in V(T)$ and an index~$j \in [r-1]$ such that~$\{v_j\} \cup X \subseteq \X_{i_1}$. As~$\{v_{r+1}\}\cup (N_{G'}(v_{r+1})\cap X)$ is a clique in~$G'$, there is a bag~$i_2$ such that~$\X_{i_2}$ contains~$\{v_{r+1}\}\cup (N_{G'}(v_{r+1})\cap X)$. From the definition of tree decomposition, and as~$\{v_{r-1},v_{r+1}\}$ is an edge in~$G'$, it follows that there must be a bag~$i_3$ with~$\{v_{r-1}, v_{r+1}\} \in \X_{i_3}$.

At least one of the following cases must hold: \begin{enumerate}
\item Node~$i_3$ is on the path in~$T$ from~$i_1$ to~$i_2$.
\item Node~$i_1$ is on the path in~$T$ from~$i_2$ to~$i_3$.
\item Node~$i_2$ is on the path in~$T$ from~$i_1$ to~$i_3$.
\item There is a node~$i_4$ such that~$i_1$,~$i_2$, and~$i_3$ are in
different subtrees of~$T-\{i_4\}$.
\end{enumerate}

\noindent (The order of the cases is chosen in this way to make the proof easier to follow.)

\paragraph{Case 1:~$i_3$ is on the path in~$T$ from~$i_1$ to~$i_2$.}
Note that~$N_G(v_r)\cap X \subseteq N_{G'}(v_{r+1})\cap X \subseteq \X_{i_1} \cap \X_{i_2}$. So~$N_G(v_r)\cap X \subseteq \X_{i_3}$. Now add a new bag~$i'$ to the tree decomposition of~$G'$. Make~$i'$ adjacent to~$i_3$, and set~$\X_{i'} := \{v_{r-1}, v_r, v_{r+1}\} \cup (N_G(v_r)\cap X)$. This is a tree decomposition of~$G$. The new bag has size at most~$3+|X| \leq k+1$ so the width is at most~$k$, which concludes this case.

\bigskip
This first case was convenient since it guaranteed the existence of a bag containing~$N_G(v_r)$, making it possible to construct a tree decomposition for the original graph~$G$ by just appending another bag there containing~$N_G(v_r) \cup \{v_r\}$. In the remaining cases we will have to restructure the tree decomposition more severely. The main idea behind each of the remaining cases will be to find a suffix $(v_{j'}, v_{j'+1}, \ldots, v_{r-1})$ of the clique-seeing path such that there exists a special bag containing~$v_{j'}, v_{r+1}$, and all the neighbors that the vertices of~$\{v_{j'}, \ldots, v_{r-1}\}$ have in the set~$X$. We then build a new tree decomposition by taking the vertices of the path suffix out of the existing bags, making a path decomposition for this suffix to which we add all the path's neighbors in~$X$, and attaching this path decomposition to the special bag. So let us now go into details.

\paragraph{Case 2:~$i_1$ is on the path in~$T$ from~$i_2$ to~$i_3$.} 
From the definition of a tree decomposition, it follows that~$v_{r+1}\in \X_{i_1}$. We now modify the tree decomposition as follows:
\begin{itemize}
\item For~$j'$ with~$j < j' \leq r-1$, remove~$v_{j'}$ from all (old) bags that it appears in. (To clarify, if~$j=r-1$ then we do not remove any vertices.)
\item Add new bags~$i'_j$,~$i'_{j+1}, \ldots, i'_{r-1}$.
\item Make bag~$i'_j$ adjacent in~$T$ to~$i_1$.
\item For~$j'$, with~$j\leq j' < r-1$, make bag~$i'_{j'}$ adjacent to bag~$i'_{j'+1}$.
\item For~$j'$, with~$j\leq j' \leq r-1$, set~$\X_{i_{j'}} := X \cup \{v_{j'}, v_{j'+1}, v_{r+1}\}$.
\end{itemize}

\noindent One can verify that this is indeed a tree decomposition of~$G$. (E.g., the bag~$i'_{r-1}$ contains~$v_r$ and all its neighbors.) As the new bags have size bounded by~$|X|+3\leq k+1$, the width of the tree decomposition is at most~$k$.

\paragraph{Case 3:~$i_2$ is on the path in~$T$ from~$i_1$ to~$i_3$.}
Consider the path from~$v_j$ to~$v_{r-1}$. As~$v_j\in \X_{i_1}$ and~$v_{r-1}\in \X_{i_3}$, by \lemmaref{lemma:pathcontainment} there is a~$j'$ with~$j\leq j'\leq r-1$ such that~$v_{j'}\in \X_{i_2}$. Consider the largest such index~$j' := \max \{ j^* \mid j \leq j^* \leq r-1 \wedge v_{j^*} \in \X_{i_2} \}$.

Consider a value~$j''$ with~$j'<j''\leq r-1$; we will show that any neighbors of~$v_{j''}$ in the set~$X$ must be contained in~$\X_{i_2}$. To this end, consider a vertex~$w\in N_G(v_{j''})\cap X$. There must be some bag~$\X_{i_0}$ containing both~$v_{j''}$ and~$w$. Node~$i_0$ must belong to the same subtree of~$T - \{i_2\}$ as~$i_3$: if this is not the case, then~$i_2$ lies on the path from~$i_0$ to~$i_3$ which implies by \lemmaref{lemma:pathcontainment} and the existence of the path~$(v_{j''}, v_{j''+1}, \ldots, v_{r-1})$ that~$\X_{i_2}$ contains a vertex of that path with index~$j''$ or higher; but this contradicts our choice of~$j'$. 

Hence~$i_0$ indeed belongs to the same subtree as~$i_3$. Since~$w\in X\subseteq \X_{i_1}$ and~$w \in \X_{i_0}$, the properties of tree decompositions now ensure that~$w\in \X_{i_2}$.  As~$w$ and~$j''$ were arbitrary this shows that~$\bigcup_{j'<j''\leq r-1} N_G(v_{j''})\cap X \subseteq \X_{i_2}$.

Write~$Z= \bigcup_{j'<j''\leq r} N_G(v_{j''})\cap X$, and observe carefully that in the definition of~$Z$ the union also includes~$v_r$, which was excluded in the earlier formula. As~$N_G(v_{r}) \cap X \subseteq N_{G'}(v_{r+1}) \cap X$ (as we obtained~$v_{r+1}$ from a contraction that involved~$v_r$), we have that~$Z\subseteq \X_{i_2}$. We now modify the tree decomposition, more or less similarly as in the previous case:
\begin{itemize}
	\item For~$j''$, with~$j' < j'' \leq r-1$, remove~$v_{j''}$
from all (old) bags that it appears in.
	\item Add new bags~$i'_{j'}$,~$i'_{j'+1}, \ldots, i'_{r-1}$.
	\item Make bag~$i'_{j'}$ adjacent in~$T$ to~$i_2$.
	\item For~$j''$ with~$j'\leq j'' < r-1$, make bag~$i'_{j''}$ adjacent to bag~$i'_{j''+1}$.
	\item For~$j''$ with~$j'\leq j'' \leq r-1$, set~$\X_{i_{j''}} := Z \cup \{v_{j''}, v_{j''+1}, v_{r+1}\}$.
\end{itemize}

\noindent As in the previous case, this is a tree decomposition of width at most~$k$ of~$G$.

\paragraph{Case 4: There is a node~$i_4$ such that~$i_1$,~$i_2$, and~$i_3$ belong to different subtrees of~$T-\{i_4\}$.}
The analysis is more or less similar as in Case 3. We have that~$v_{r+1}\in \X_{i_4}$, and~$N_G(v_{r}) \cap X \subseteq \X_{i_1} \cap \X_{i_2}$, hence~$N_G(v_r) \cap X \subseteq \X_{i_4}$.

We have~$v_j\in \X_{i_1}$ and~$v_{r-1}\in \X_{i_3}$, so by \lemmaref{lemma:pathcontainment} there is a~$j'$ with~$j\leq j'\leq r-1$ such that~$v_{j'}\in \X_{i_4}$. Assume~$j'$ is the maximum value with~$j\leq j'\leq r-1$ and~$v_{j'}\in \X_{i_4}$. 

Consider a~$j''$ with~$j'< j'' \leq r-1$. Since $v_{r-1} \in \X_{i_3}$, all bags containing~$v_{j''}$ must belong to the same subtree of~$T-\{i_4\}$ as~$i_3$, as otherwise~$i_4$ must contain a vertex from~$\{v_{j'+1}, v_{j'+2}, \ldots, v_{r-1}\}$. All neighbors of~$v_{j''}$ in~$X$ belong to some bag in the same subtree of~$T-\{i_4\}$ as~$i_2$. As~$N_G(v_{j''}) \cap X \subseteq X \subseteq \X_{i_1}$, we have that~$N_G(v_{j''}) \cap X \subseteq \X_{i_4}$. We can now use the same construction as in Case 3, except that we attach the new part of the tree decomposition to~$i_4$.

Thus, in all cases, we obtain a tree decomposition of~$G$ of width at most~$k$, and conclude that the rule is safe.
\end{proof}

\subsubsection{Correctness of Rule \ref{rule:cliqueseeingsep}}

This section is devoted to the correctness proof of \ruleref{rule:cliqueseeingsep}. The following lemma will be useful when giving the correctness proof.

\begin{lemma} \label{lemma:treewidth:max:separated:vertex}
Let~$G$ be a graph, let~$Y \subseteq V(G)$, and let~$v^* \not \in Y$ be a distinguished vertex in~$G$. If, for each~$u \not \in Y \cup \{v^*\}$, there is a clique separator~$Z \subseteq Y$ that separates~$u$ from~$v^*$, then the treewidth of~$G$ is the maximum of~$\tw(G - \{v^*\})$ and~$\tw(G[Y \cup \{v^*\}])$.
\end{lemma}
\begin{proof}
As treewidth does not increase when taking subgraphs, it is obvious that~$\tw(G)$ is at least the maximum. Define~$R := V(G) \setminus (Y \cup \{v^*\})$. Using induction on~$|R|$ we prove that~$\tw(G)$ is at most the maximum. If~$R = \emptyset$ then~$G = G[Y \cup \{v^*\}]$ so the claim is trivial. Otherwise, let~$u \in R$ and let~$Z \subseteq Y$ be a clique separator that separates~$u$ from~$v^*$. Let~$V_1, \ldots, V_r$ be the connected components of~$G - Z$, and assume without loss of generality that~$v^* \in V_1$. By \lemmaref{lemma:cliqueseparator} the treewidth of~$G$ equals the maximum of~$\tw(G[Z \cup V_i])$ over all relevant choices of~$i$. We distinguish two cases.

First assume that there is an index~$i > 1$ such that~$\tw(G) = \tw(G[Z \cup V_i])$. Observe that~$G[Z \cup V_i]$ is a subgraph of~$G - \{v^*\}$, as~$v^* \in V_1$. This implies that~$\tw(G - \{v^*\}) \geq \tw(G[Z \cup V_i]) = \tw(G)$. Hence the treewidth of~$G$ is at most the claimed maximum.

In the second case, assume that~$i=1$ is the unique index for which~$\tw(G)$ equals~$\tw(G[Z \cup V_i])$. We invoke the induction hypothesis to complete the proof. Let~$G' := G[Z \cup V_1]$, and let~$Y' := Y \cap V(G')$. Define~$R' := V(G') \setminus (Y' \cup \{v^*\})$, which is a proper subset of~$R$ as it does not contain the vertex~$u$ that was separated from~$v^*$ by~$Z$. By the case distinction we have~$\tw(G) = \tw(G')$. For each vertex~$u' \in R'$, there is a clique separator~$Z \subseteq Y$ separating~$u'$ from~$v^*$ in the graph~$G$, by assumption. Putting~$Z' := Z \cap V(G')$ one may easily verify that~$Z' \subseteq Y'$ is a clique in~$G'$ that separates~$u'$ from~$v^*$. Hence all preconditions of the induction hypothesis are satisfied. As~$|R'| < |R|$ we may apply induction to~$G'$, $Y'$, and~$v^*$ to establish that:
$$\tw(G') = \tw(G) = \max (\tw(G' - \{v^*\}), \tw(G'[Y' \cup \{v^*\}])).$$
Now compare the two terms in this maximum to those in the statement of the claim. By the subgraph relation we have~$\tw(G - \{v^*\}) \geq \tw(G' - \{v^*\})$ and~$\tw(G[Y \cup \{v^*\}]) \geq \tw(G'[Y' \cup \{v^*\}])$. Hence we find:
$$\max (\tw(G - \{v^*\}), \tw(G[Y \cup \{v^*\}])) \geq \max (\tw(G' - \{v^*\}), \tw(G'[Y' \cup \{v^*\}])) = \tw(G).$$
This proves the claim.
\end{proof}

Using this lemma we establish the correctness of \ruleref{rule:cliqueseeingsep}.

\begin{lemma}\label{lemma:rule:cliqueseeingsep:safe}
\ruleref{rule:cliqueseeingsep} is safe.
\end{lemma}

\begin{proof}
Let us first recall the statement of \ruleref{rule:cliqueseeingsep}. It requires that we have a clique-seeing path $(v_0,v_1, v_2, v_3, v_4)$ in~$G$, whose vertices avoid~$S$, that sees a clique~$X\subseteq S$. Furthermore, it requires that $\{v_1,v_2,v_3\}\cup X$ separates~$v_0$ from~$v_4$ and that \ruleref{rule:almostcliqueseparator} cannot be applied. Then, we compute the treewidth of~$G[\{v_1,v_2,v_3\} \cup X]$; if that is larger than~$k$ we answer~$\no$, else we remove~$v_2$ from~$G$. To prove the correctness of the rule, we establish a claim about the structure of the instance.

\begin{claim}
For each vertex~$u \not \in \{v_1,v_2,v_3\} \cup X$ there is a clique separator $Z \subseteq \{v_1, v_3\} \cup X$ that separates~$u$ from~$v_2$.
\end{claim}
\begin{proof}
We first identify a number of vertex sets.
\begin{itemize}
\item[$A$:] the vertices in the same connected component as~$v_0$ in the graph~$G - (\{v_1\}\cup (N_G(v_1)\cap X))$.
\item[$B$:] the vertices in the same connected component as~$v_4$ in the graph~$G - (\{v_3\}\cup (N_G(v_3)\cap X))$.
\item[$C$:] all vertices \emph{not} in the same connected component as~$v_2$ in the graph~$G-X$.
\end{itemize}

We prove that each vertex~$u \not \in \{v_1,v_2,v_3\} \cup X$ is contained in at least one of~$A,B,C$. We do this by showing that all such vertices~$u$ which are not contained in~$C$, must be contained in~$A$ or~$B$. So assume there is some vertex~$u$ which is not in set~$C$ because there exists a path~$P$ from~$u$ to~$v_2$ in~$G - X$. The first vertex of the clique-seeing path visited by~$P$ when starting in~$u$, is either~$v_0$ or~$v_4$; there is no other way to enter the clique-seeing path. Hence~$G-X$ has a subpath~$P' \subseteq P$ from~$u$ to either~$v_0$ or~$v_4$, whose internal vertices do not belong to the clique-seeing path. As this path avoids~$\{v_1, v_2, v_3\} \cup X$, which is a superset of the separators used when defining~$A$ and~$B$, this path witnesses the fact that~$u$ belongs to~$A$ or~$B$.

Now we are in position to prove the claim. It is easy to see that the sets~$\{v_1\} \cup (N_G(v_1) \cap X)$, $\{v_3\} \cup (N_G(v_3) \cap X)$, and~$X$, are all cliques, and subsets of $\{v_1, v_3\} \cup X$. As the preconditions to the reduction rule contain all the conditions of \lemmaref{lemma:separatorsOnCliqueSeeingPath}, that lemma shows that the set~$\{v_1\} \cup (N_G(v_1) \cap X)$ separates~$v_0$ from~$v_2$. Consequently, it separates all vertices in~$A$ from~$v_2$. Another application of the lemma shows that~$\{v_3\} \cup (N_G(v_3) \cap X)$ separates~$v_2$ from~$v_4$, and therefore separates~$v_2$ from~$B$. Finally, the set~$X$ separates all vertices of~$C$ from~$v_2$; this completes the proof of the claim.
\end{proof}

Using the claim we can give the correctness argument for the reduction rule. Apply \lemmaref{lemma:treewidth:max:separated:vertex} to establish that the treewidth of~$G$ is the maximum of~$\tw(G - \{v_2\})$ and~$\tw(G[\{v_1,v_2,v_3\} \cup X])$: use~$X \cup \{v_1, v_3\}$ as the set~$Y$, vertex~$v_2$ as the special vertex~$v^*$, and note that the given claim proves the existence of the required clique separators. Clearly, if the treewidth of~$G[\{v_1,v_2,v_3\} \cup X]$ is greater than~$k$ it is correct to answer \no. If its treewidth is at most~$k$, then the treewidth of~$G$ is at most~$k$ if and only if the treewidth of~$G-\{v_2\}$ is at most~$k$. Hence the rule is safe.
\end{proof}

\subsection{A cut-off rule} \label{subsection:cutoffrule}
As we will show, the rules presented so far shrink any \yes-instance to~$\Oh(\ell^4)$ vertices. The final rule needed to obtain the desired kernel is therefore the following cut-off rule. It rejects instances that have long clique-seeing paths despite the exhaustive application of the previous rules.

\begin{rul}\label{rule:finalno}
%It assumes that we have a clique-seeing path~$(v_0, v_1, \ldots, v_r, v_{r+1})$ in~$G$, whose vertices avoid~$S$, that sees a clique~$X \subseteq S$. 
Suppose we have a clique-seeing path~$(v_0, \ldots, v_{r+1})$ in~$G$, whose vertices avoid~$S$, that sees a clique~$X \subseteq S$. If~$r \geq 6k+6$ and \rulesref{rule:neighborimprovementR}--\ref{rule:cliqueseeingsep} are not applicable, then answer \no.
\end{rul}

\noindent A detailed counting argument shows the safeness of this rule. We will bound the length of clique-seeing paths in terms of the size of the corresponding clique seen by the path. If such a path is too long (in comparison to its clique), then we can invoke \lemmaref{lemma:largeX} to infer the existence of a~$k+2$-clique minor (which permits us to correctly answer \no).

\begin{lemma}\label{lemma:rule:finalno:correctness123}
Suppose that none of \rulesref{rule:neighborimprovementR}--\ref{rule:cliqueseeingsep} can be applied. Let~$(v_0, v_1, \ldots, v_{r+1})$ be a clique-seeing path in~$G$, whose vertices avoid~$S$, seeing a clique~$X \subseteq S$.
\begin{enumerate}
\item \label{case:k-2}If~$|X|\leq k-2$ then~$r\leq|X|+1\leq k-1$.
\item \label{case:k-1}If~$|X|=k-1$ and~$r\geq 3k+3$ then the treewidth of~$G$ is greater than~$k$.
\item \label{case:k}If~$|X|=k$ and~$r\geq 6k+6$ then the treewidth of~$G$ is greater than~$k$.
\item \label{case:k+1}If~$|X|\geq k+1$ then the treewidth of~$G$ is greater than~$k$.
\end{enumerate}
\end{lemma}

\begin{proof}
We consider the cases consecutively.
\begin{enumerate}
\item Suppose~$|X|\leq k-2$. In particular, \ruleref{rule:subsetpath} cannot be applied. Thus for each~$i$ with~$1< i \leq r$:
\begin{equation} \nonumber
N_G(v_i) \cap X \not\subseteq \bigcup_{j=1}^{i-1} N_G(v_j)\cap X.
\end{equation}
Hence, via induction it follows that~$|\bigcup_{j=1}^{i} N_G(v_j)\cap X| \geq i+1$, and therefore that~$r \leq |\bigcup_{j=1}^{r} N_G(v_j)\cap X| \leq |X|+1$.
\item Suppose~$|X|=k-1$ and~$r\geq 3k+3$. Let 
\begin{equation} \nonumber
X_1 = \bigcup_{j=1}^{k+1} N_G(v_j)\cap X, \quad X_2 = \bigcup_{j=k+2}^{2k+2} N_G(v_j)\cap X,\quad X_3 = \bigcup_{j=2k+3}^{3k+3} N_G(v_j)\cap X.
\end{equation}

If~$X_1\neq X$, then~$(v_0, v_1, \ldots, v_{k+1}, v_{k+2})$ is a clique-seeing path with~$k+1$ internal vertices and seeing clique~$X_1$, with~$|X_1|<k-1$, contradicting \caseref{case:k-2}. Similarly, we get a contradiction when~$X_2\neq X$ or~$X_3\neq X$.

Therefore, assume~$X_1=X_2=X_3=X$. Consider the path~$(v_0,\ldots,v_{2k+3})$ seeing clique~$X$. By \proposref{proposition:connectedness} there is a path from~$v_0$ to~$v_{2k+3}$ that avoids~$\{v_1,\ldots,v_{2k+2}\}\cup X$. Hence, the component~$Z$ containing~$v_0$ in~$G-(\{v_1,\ldots,v_{2k+2}\}\cup X)$ also contains~$v_{2k+3}$, and hence~$\{v_{2k+3}, \ldots, v_{3k+3}\} \subseteq Z$. Thus by~$X_1=X_2=X_3=X$ each vertex of~$X$ has a neighbor in each of~$\{v_1,\ldots,v_{k+1}\}$,~$\{v_{k+2},\ldots,v_{2k+2}\}$, and~$\{v_{2k+3}, \ldots, v_{3k+3}\}
\subseteq Z$. Therefore, the conditions for the last case of \lemmaref{lemma:largeX} are fulfilled for~$(v_0,\ldots,v_{2k+3})$ and~$\ell=k+1$. It follows that~$G$ has treewidth at least~$k+1$.
\item Suppose~$|X|=k$ and~$r\geq 6k+6$. Let
\begin{equation} \nonumber
X_1 = \bigcup_{j=1}^{3k+3} N_G(v_j)\cap X \mbox{ and } X_2 = \bigcup_{j=3k+4}^{6k+6} N_G(v_j)\cap X. 
\end{equation}

If~$X_1\neq X$, then considering the clique-seeing path~$(v_0, v_1, \ldots, v_{3k+3}, v_{3k+4})$ we get a contradiction to \caseref{case:k-2} if~$|X_1|\leq k-2$ or the claim follows from \caseref{case:k-1} if~$|X_1|=k-1$. This can be argued similarly when~$X_2\neq X$.

Therefore,~$X_1=X_2=X$. Now consider the path~$(v_0,\ldots,v_{3k+4})$. By \proposref{proposition:connectedness} there is a path from~$v_0$ to~$v_{3k+4}$ in~$G-(\{v_1,\ldots,v_{3k+3}\}\cup X)$. It follows that the connected component~$Z$ of~$v_0$ in that graph also contains the vertices~$v_{3k+4},\ldots,v_{6k+6}$. Thus the path~$(v_0,\ldots,v_{3k+4})$ fulfills the conditions of the second case of \lemmaref{lemma:largeX}, and we conclude that the treewidth of~$G$ is at least~$k+1$.
\item This follows directly from the first case of \lemmaref{lemma:largeX}.\qedhere
\end{enumerate}
\end{proof}

Correctness of \ruleref{rule:finalno} follows immediately from the lemma: if the clique-seeing path has at least~$6k+6$ internal vertices, then (assuming that none of \rulesref{rule:neighborimprovementR}--\ref{rule:cliqueseeingsep} can be applied) it sees a clique of size at least~$k-1$ (by \caseref{case:k-2}). \casesref{case:k-1}--\ref{case:k+1} then show that answering \no is correct.

\subsection{The kernelization}
\label{section:counting}

Now we are ready to present the kernelization algorithm. It consists of the exhaustive application of \rulesref{rule:neighborimprovementR} through~\ref{rule:finalno}.

\begin{theorem}\label{theorem:fvskernel}
\treewidthByFVS has a kernel with~$\Oh(\ell^4)$ vertices.
\end{theorem}

\begin{proof}
Let~$(G,k,S)$ be an instance of \treewidthByFVS and let~$(G',k,S')$ be obtained from exhaustive application of \rulesref{rule:neighborimprovementR} through~\ref{rule:finalno}. Observe that \rulesref{rule:subsetpath}--\ref{rule:finalno} are only tested for paths in~$G-S$; as that is a forest, we need to test at most~$\Oh(n^2)$ paths. Thus the test has to be done a polynomial number of times. We showed that all rules are safe and that they can be performed exhaustively in polynomial time. Hence the instances are equivalent,~$S'$ is a feedback vertex set of~$G'$, and~$|S'|\leq|S| = \ell$.

Let us analyze the size of~$G'$. The forest~$G'-S'$ has~$\Oh(\ell^3)$ leaves by \proposref{proposition:tja}, and hence~$\Oh(\ell^3)$ vertices of degree at least three. There are~$\Oh(\ell^3)$ paths in the forest whose endpoints are leaves or vertices of degree at least three and whose internal vertices have degree two each. Each path of length at least~$6k+8$, i.e., with at least~$6k+6$ internal vertices is not clique-seeing by \ruleref{rule:finalno}. Thus, for the analysis, we split the paths into parts of size~$6k+8$ which are therefore not clique-seeing. At most~$6k+7$ vertices per path will not belong to such a part, but these are at most~$\Oh(\ell^3\cdot k)$ in total. We assign each part to a pair of non-adjacent vertices in~$S$ which are adjacent to internal vertices of the part. We can assign at most~$k$ parts to a pair~$\{u,v\}$: indeed, since the parts are disjoint they would otherwise give rise to more than~$k$ disjoint~$u-v$ paths,  contradicting~$(G',k,S')$ being reduced under \ruleref{rule:neighborimprovementR}. Thus, we have~$\Oh(k(6k+8)\ell^2)+\Oh(\ell^3\cdot k)=\Oh(\ell^4)$ vertices in the forest~$G'-S'$, and thus~$\Oh(\ell^4)$ vertices in~$G'$.
\end{proof}

Using a~$2$-approximation algorithm for feedback vertex set we obtain a corollary similar to the one given in the previous section.

\begin{corollary}
There is a polynomial-time algorithm that given an instance $(G,k)$ of \treewidth computes an equivalent instance $(G',k)$ such that~$V(G') \subseteq V(G)$ and~$|V(G')| \in \Oh(\fvs(G)^4)$, where~$\fvs(G)$ is the size of a minimum feedback vertex set of~$G$.
\end{corollary}

\section{Kernelization lower bounds}\label{section:lowerbounds}
In this section we present our lower bound results for \treewidth and \weightedtreewidth. We define the lower bound framework in \sectref{sect:tw:lowerboundtools}, and prove the lower bounds in the subsequent sections. For completeness we show in \sectref{sect:tw:fptClassifications} that the considered problems are fixed-parameter tractable, and therefore admit (superpolynomial) kernelizations.

\subsection{Lower Bound Tools} \label{sect:tw:lowerboundtools}
We use the notion of a cross-composition~\cite{BodlaenderJK11} which is an extension of the known lower bound techniques based on work of Bodlaender et al.~\cite{BodlaenderDFH09} as well as of Fortnow and Santhanam~\cite{FortnowS11}. We first give two necessary definitions.

\begin{definition}[\cite{BodlaenderJK11}]%[Polynomial equivalence relation] 
\label{polyEquivalenceRelation}
An equivalence relation~\R on~$\Sigma^*$ is called a \emph{polynomial equivalence relation} if the following two conditions hold:
\begin{enumerate}
	\item There is an algorithm that given two strings~$x,y \in \Sigma^*$ decides whether~$x$ and~$y$ belong to the same equivalence class in~$(|x| + |y|)^{\Oh(1)}$ time.
	\item For any finite set~$S \subseteq \Sigma^*$ the equivalence relation~\R partitions the elements of~$S$ into at most~$(\max _{x \in S} |x|)^{\Oh(1)}$ classes.
\end{enumerate}
\end{definition}

\begin{definition}[\cite{BodlaenderJK11}]%[Cross-composition] 
\label{crossComposition}
Let~$L \subseteq \Sigma^*$ be a set and let~$Q \subseteq \Sigma^* \times \mathbb{N}$ be a parameterized problem. We say that~$L$ \emph{cross-composes} into~$Q$ if there is a polynomial equivalence relation~\R and an algorithm which, given~$t$ strings~$x_1, x_2, \ldots, x_t$ belonging to the same equivalence class of~\R, computes an instance~$(x^*,k^*) \in \Sigma^* \times \mathbb{N}$ in time polynomial in~$\sum _{i\in[t]} |x_i|$ such that:
\begin{enumerate}
	\item $(x^*, k^*) \in Q \Leftrightarrow x_i \in L$ for some~$i \in [t]$,
	\item $k^*$ is bounded by a polynomial in~$\max _{i\in[t]} |x_i|+\log t$.
\end{enumerate}
\end{definition}

It is known that a cross-composition of any NP-hard set into a parameterized problem~$Q$ rules out polynomial kernels for~$Q$ \cite{FortnowS11,BodlaenderDFH09,BodlaenderJK11}.

\begin{theorem}[\cite{BodlaenderJK11}, Corollary 10] \label{crossCompositionNoKernel}
If some set~$L$ is NP-hard under Karp reductions and~$L$ cross-composes into the parameterized problem~$Q$ then there is no polynomial kernel for~$Q$ unless \containment.
\end{theorem}

\subsection{A kernelization lower bound for Treewidth parameterized by distance from a single Clique} \label{sect:tw:distanceClique}
In this section we show that \treewidthByClique (\TWMSC) does not admit a polynomial kernel unless \containment. The problem is covered by the general template given in the introduction, when using~$\F$ as the class of all cliques. Observe carefully that~$\F$ only contains \emph{connected} graphs, and in particular~$\F$ is not closed under disjoint union.

To prove the lower bound we employ the technique of cross-composition: we show that \textsc{Cutwidth on subcubic graphs} cross-composes into \TWMSC. Since the cutwidth problem is still NP-complete under the stated restriction~\cite[Corollary 2.10]{MonienS88}, this will give a lower bound using \thmref{crossCompositionNoKernel}. The cutwidth problem we use is formally defined as follows:

\problemdef
{Cutwidth on subcubic graphs (\cutwidththree)}
{A graph~$G$ on~$n$ vertices in which each vertex has degree at most three, and an integer~$k \leq |E(G)|$.}
{Is there a linear layout of~$G$ of cutwidth at most~$k$, i.e., a permutation~$\pi \colon V(G) \to [n]$ of~$V(G)$ such that~$\max _{i\in[n]} | \{ \{u,v\} \in E(G) \mid \pi(u) \leq i < \pi(v) \}| \leq k$?}

To obtain a kernel lower bound through cross-composition, we have to embed the logical OR of a series of~$t$ input instances of \cutwidththree on~$n$ vertices each into a single instance of the target problem for a parameter value polynomial in~$n + \log t$. At the heart of our construction lies an idea of Arnborg et al.~\cite{ArnborgCP87} employed in their NP-completeness proof for \treewidth. They interpreted the treewidth of a graph as the minimum cost of an elimination ordering on its vertices, and showed how for a given graph~$G$ a co-bipartite graph~$G^*$ can be created such that the cost of elimination orderings on~$G^*$ corresponds to the cutwidth of~$G$ under a related ordering. Our composition extends their construction significantly. We express it in terms of weighted treewidth, and afterwards use a simple transformation to reduce it to the unweighted setting.

Before giving the proof, we sketch the main ideas. By the degree bound, instances with~$n$ vertices have~$\Oh(n^2)$ different degree sequences. The framework of cross-composition thus allows us to work on instances with the same degree sequence (and same~$k$). By enforcing that the structure of one side of the co-bipartite graph~$G^*$ only has to depend on this sequence, all inputs can share the same ``right hand side'' of the co-bipartite graph; this part will remain small and act as the modulator. By a careful balancing act of weight values we then enforce that the cost of elimination orderings on the constructed graph~$G^*$ are dominated by eliminating the vertices corresponding to exactly one of the input instances, ensuring that a sufficiently low treewidth is already achieved when the answer to one of the input instances is \yes. On the other hand, the use of a binary-encoding representation of instance numbers ensures that low-cost elimination orderings for~$G^*$ do not mix vertices corresponding to different input instances. Before presenting the lower bound, we introduce some terminology and results for \weightedtreewidth.

\begin{definition}
Consider a graph~$G$ weighted by function~$w$ along with an elimination ordering~$\pi$ on the vertices of~$G$. Eliminate the vertices of~$G$ in the order given by~$\pi$. Then the \emph{cost} of~$\pi$ is the maximum over all vertices~$v \in V(G)$ of~$\sum _{u \in N[v]} w(u)$ at the time that~$v$ is eliminated.
\end{definition}

The following lemma is a weighted analogue of a result by Bodlaender et al.~\cite{BodlaenderFKKT12a}. They showed that for any clique~$B$ in a graph~$G$, there is an optimal elimination ordering that ends with~$B$. The lemma will be useful when proving the correctness of our cross-composition, because it allows us to assert the existence of well-behaved optimal elimination orderings for the constructed instance. A related statement in the restricted setting of co-bipartite graphs was already employed by Arnborg et al.~\cite{ArnborgCP87} in their NP-completeness proof.

\begin{lemma}[Cf. {\cite[Lemma 4.3]{BodlaenderFKKT12a}}] \label{eliminatePartiteSetFirst}
Let~$G$ be a graph weighted by function~$w$ containing a clique~$B \subseteq V(G)$, and let~$A := V(G) \setminus B$. There is a minimum-cost elimination ordering~$\pi^*$ of~$G$ that eliminates all vertices of~$A$ before eliminating any vertex of~$B$.
\end{lemma}
\begin{proof}
Let~$\pi$ be a minimum-cost elimination ordering of~$G$ with cost~$k$. The filled graph of~$G$ with respect to~$\pi$ is a chordal supergraph~$H$ of~$G$, as described in \sectref{section:treewidthdef}. We claim that the maximum weight of a clique in~$H$ is bounded by~$k$. To see this, assume that~$H$ contains a clique~$C \subseteq V(H)$ of weight more than~$k$. Let~$v$ be the first vertex of the clique that is eliminated by ordering~$\pi$. Then the ordering incurs cost~$\sum _{u \in C} w(u)$ when eliminating~$v$: the edges between~$v$ and the other members of~$C$ must already have been filled at that point, since they cannot become part of the filled graph after~$v$ is eliminated. Hence the existence of a clique of weight more than~$k$ would imply that~$\pi$ has cost more than~$k$, a contradiction.

Construct a perfect elimination ordering~$\pi'$ of~$H$ as follows. As long as~$A$ is nonempty, there must be a vertex in~$A$ that is simplicial in~$H$: if~$H$ is a clique then any vertex of~$A$ is simplicial, and if~$H$ is not a clique then by \proposref{proposition:nonAdjacentSimplicial} there are two nonadjacent simplicial vertices in~$H$. These cannot both belong to~$B$, since~$B$ is a clique; hence at least one belongs to~$A$. We pick such a simplicial vertex in~$A$, remove it from the graph, and repeat. Once~$A$ is empty, end with the vertices in the clique~$B$ in arbitrary order; they are all simplicial. We end up with a perfect elimination ordering~$\pi'$ of~$H$: since we only pick vertices that are simplicial at the time they are picked, they produce no fill edges in~$H$. This property implies that when a vertex is eliminated under~$\pi'$ in~$H$, its closed neighborhood forms a clique in~$H$ and therefore has weight at most~$k$.

Now consider the cost of using ordering~$\pi'$ on the weighted graph~$G$. Since~$G$ is a subgraph of~$H$, for each vertex~$v$ its neighborhood at the time of elimination is a subset of what would have been its neighborhood on elimination from the graph~$H$. Since the closed neighborhood weight in~$H$ is at most~$k$, the weight of the neighborhood when eliminating it from~$G$ cannot be larger. Hence~$\pi'$ is an elimination ordering of~$G$ with cost at most~$k$ under~$w$. As it eliminates all vertices of~$A$ before eliminating any vertex in~$B$, this concludes the proof.
\end{proof}

The following proposition is the weighted variant of the folklore fact that treewidth can be characterized by elimination orderings. For completeness we sketch the proof in the weighted setting.

% Use {\cite[Theorem 8]{BodlaenderR03}} somewhere? They say weighted tw == max weight of clique in chordal supergraph.
\begin{proposition} \label{eliminationCostIsWeightedWidth}
Graph~$G$ has weighted treewidth at most~$k$ if and only if there is an elimination ordering of~$G$ of cost at most~$k+1$.
\end{proposition}
\begin{proof}
In one direction, let~$G$ have a tree decomposition of weighted width at most~$k$. By completing the vertices of each bag into a clique we obtain a chordal graph~$H$ whose maximal cliques correspond to the bags of the decomposition. As the maximum weight of a bag is at most~$k+1$, the cliques in this chordal graph have weight at most~$k+1$. Now construct a perfect elimination ordering of~$G$ by repeatedly picking simplicial vertices and removing them from the graph; this ordering has cost at most~$k+1$ since for each vertex, its closed neighborhood when it is eliminated forms a clique in~$H$.

In the reverse direction, suppose that~$G$ has an elimination ordering~$\pi$ of cost at most~$k+1$. The filled chordal supergraph corresponding to the ordering has no clique of weight exceeding~$k+1$, as otherwise the vertex in this clique that is eliminated first would cause the cost to exceed~$k+1$ as in the proof of \lemmaref{eliminatePartiteSetFirst}. As any chordal graph has a tree decomposition whose bags are the maximal cliques, no bag in such a tree decomposition has weight exceeding~$k+1$. Consequently such a decomposition has weighted width at most~$k$, proving the claim.
\end{proof}

By considering the effect of updating an ordering such that it eliminates vertices with smaller neighborhoods earlier, we may observe the following.

\begin{observation} \label{eliminateLargerNeighborhoodLater}
Let~$G$ be a graph with weight function~$w$ containing two adjacent vertices~$u,v$ such that~$N_G[u] \subseteq N_G[v]$. Let~$\pi$ be an elimination ordering of~$G$ that eliminates~$v$ before~$u$, and let the ordering~$\pi'$ be obtained by updating~$\pi$ such that it eliminates~$u$ just before~$v$. Then the cost of~$\pi'$ is not higher than the cost of~$\pi$.
\end{observation}

Finally, we will need the following transformation to reduce the weighted problem to the unweighted setting.

\begin{proposition} \label{treewidthVersusWeightedWidth} % TODO: Update remainder for the fact that we changed the definition of weighted treewith by -1, thus the minus one disappeared from this proposition.
Let~$G$ be a graph with positive integral vertex weights given by function~$w$. Let~$G'$ be the graph obtained from~$G$ by iterating the following procedure: as long as there is a vertex~$v$ with weight more than one, subtract one from the weight of~$v$ and add a new vertex~$v'$ of weight one and with open neighborhood~$N[v]$. Then the treewidth of~$G'$ equals the weighted treewidth of~$G$ under~$w$.
\end{proposition}
\begin{proof}
Given a tree decomposition of~$G$, replace the weighted vertices in each bag by the unweighted copies. A bag of total weight~$k$ is transformed into a bag of size~$k$, and it is easy to verify that the decomposition satisfies all relevant conditions.

In the reverse direction, consider a tree decomposition of the unweighted graph~$G'$ of width~$k$. Make a tree decomposition of~$G$ on the same host tree, putting a weighted vertex~$v$ in a bag if all its unweighted copies are contained in that bag. As the total size of a bag in the unweighted decomposition is at most~$k+1$, the maximum weight of the corresponding bag in the weighted decomposition is at most~$k+1$. The convexity of the original decomposition ensures that all vertices induce connected subtrees. To see that each edge of the weighted graph is covered by this decomposition, note that for an edge~$\{v,w\}$ in~$G$ the unweighted vertices corresponding to~$v$ and~$w$ together form a clique in~$G'$. Hence by \proposref{proposition:simpleCliqueContainment} there is a bag containing all unweighted vertices corresponding to~$v$ or~$w$, and this bag will cover the edge~$\{v,w\}$ in the resulting decomposition of the weighted graph.
\end{proof}

% TODO: Check fix for the definition of cutwidth (there is only one \leq, not both).

\begin{theorem}\label{theorem:twmsc:nokernel}
\treewidthByClique does not admit a polynomial kernelization unless \containment.
% (which would imply a collapse of the polynomial hierarchy its third level).
\end{theorem}
\begin{proof}
We show that the NP-complete \cutwidththree problem cross-composes into \TWMSC. We start by defining a polynomial equivalence relation~$\R$. Fix an encoding of instances of \cutwidththree, and choose~$\R$ such that all strings that do not encode a valid instance are equivalent. For the strings that \emph{do} encode a valid instance, define two instances~$(G_1, k_1)$ and~$(G_2, k_2)$ to be equivalent if all of the following hold:~$k_1 = k_2$,~$|V(G_1)| = |V(G_2)|, |E(G_1)| = |E(G_2)|$, and for each integer~$i \in \{0,1,2,3\}$ the number of degree-$i$ vertices in~$G_1$ and~$G_2$ is the same. Since a set of valid instances on at most~$n$ vertices each is partitioned into at most~$n \times n \times n^2 \times n^3$ equivalence classes, this constitutes a polynomial equivalence relation.

We now show how to cross-compose a set of instances of \cutwidththree that belong to the same equivalence class of~$\R$. If all instances are malformed, then this can be recognized in polynomial time and we simply output a constant-size \no-instance. So in the remainder we may assume that all input instances~$(G_1, k_1), \linebreak[1] \ldots, (G_t, k_t)$ are well-formed and belong to the same equivalence class; in particular~$k_1 = \ldots = k_t = k$ and~$|V(G_1)| = \ldots = |V(G_t)| = n$. Order the vertices within each graph by increasing degree, breaking ties arbitrarily. The choice of~$\R$, together with the fact that each~$G_i$ has maximum degree three, guarantees that each graph has the same number of vertices of each degree.

Since \cutwidth on a graph on~$n$ vertices can be solved in~$\Oh^*(2^n)$ time~\cite[Theorem 10]{BodlaenderFKKT12}, we may assume that~$n \geq \log t$. For if~$n < \log t$ then applying the algorithm by Bodlaender et al.~\cite{BodlaenderFKKT12} consecutively on each instance can be done in time which is polynomial in the total input size (which is at least~$t$); we could then give a constant-size instance with the appropriate answer as the output of the cross-composition. For similar reasons we may assume~$n \geq 2$. Finally, we may assume that the number of input instances~$t$ is a power of~$2$ since we can duplicate some instances without changing the value of the OR, increasing the input size by at most a factor two.

To construct the instance of \TWMSC that encodes the OR of the input instances, we use a two-stage process for the ease of presentation. We first show that the OR of the input instances can be encoded into an instance of \weightedTreewidthByClique on a co-bipartite graph with partite sets~$A$ and~$B$, such that the total weight of the set~$B$ is polynomial in~$n + \log t$. The set~$B$ will be the modulator, which is valid since removing the partite set~$B$ from a co-bipartite graph leaves a clique. We then use \proposref{treewidthVersusWeightedWidth} to obtain an equivalent instance of \TWMSC, and since the total weight of~$B$ is sufficiently small this produces an instance of \TWMSC that encodes the OR of the input instances, and has a modulator to a single clique whose size is polynomial in~$n + \log t$.

We now construct a graph~$G^*$ and weight function~$w^*$ such that computing the weighted treewidth of~$G^*$ corresponds to computing the OR of the instances of \cutwidththree. The construction is based on the NP-completeness proof for \treewidth by Arnborg et al.~\cite{ArnborgCP87}. The graph~$G^*$ will be co-bipartite with partite sets~$A$ and~$B$, so~$V(G^*) := A \dot \cup B$ and~$A$ and~$B$ are cliques in~$G^*$. The graph~$G^*$ is defined as follows:
\begin{itemize}
	\item For each input graph~$G_i$ with~$i \in [t]$, for each vertex~$j \in V(G_i)$, we add a vertex~$v_{i,j}$ of weight~$n^3$ to~$A$ which corresponds to vertex~$j$. For a given value of~$j \in [n]$ we say that the vertices~$v_{i,j}$ (for all relevant values of~$i$) are \emph{$A$-representatives} of node~$j$. We also add a \emph{dummy} vertex~$d_i$ for each instance~$i \in [t]$ to~$A$ of weight~$n^6$. We turn~$A$ into a clique.
	\item The vertex set~$B$ consists of three parts: the \emph{instance selector vertices}~$B_I$, the \emph{node representatives}~$B_N$, and the \emph{edge representatives}~$B_E$.
	\begin{itemize}
		\item The instance selector vertices will be used to encode the binary representation of instance numbers. Since we assumed~$t$ to be a power of two, we need~$\log t$ bits to encode an instance number and therefore~$2 \log t$ vertices are used to represent all possible bit values for~$\log t$ positions. So~$B_I := \{ a_q, b_q \mid q \in [\log t] \}$. Each vertex in~$B_I$ has weight~$n^5$.
		
		We connect the vertices of~$B_I$ to the vertices of~$A$ as follows. We make a vertex~$v_{i,j}$ in~$A$ (which corresponds to instance~$i$) adjacent to the instance selector vertices of the bit values of the binary representation of~$i$. So for~$q \in [\log t]$, if the $q$-th bit of number~$i$ is~$1$ then we make~$v_{i,j}$ adjacent to~$a_q$, and if the bit is~$0$ then we make the vertex adjacent to~$b_q$. The adjacency from dummy vertices~$d_i$ to the vertices of~$B_I$ is defined exactly the same through the binary representation of~$i$.
		
		\item The node representatives~$B_N$ contain a vertex for each node number in~$[n]$. Recall that all input graphs have the same number of vertices of each degree, and that we sorted the vertices by degree. When we write~$\deg(j)$ for~$j \in [n]$ we will therefore take this to mean the value~$d$ such that in each input graph, the $j$-th vertex has degree~$d$. For each~$j \in [n]$ we add a vertex~$x_j$ to the set~$B_N$ and give it weight~$n^3 - \deg(j)$. The vertex~$x_j$ is said to be the (unique) \emph{$B$-representative} of node~$j$.
		
		The adjacency between~$B_N$ and~$A$ is simple: for each~$j \in [n]$ we make all $A$-representatives of~$j$ adjacent to the single $B$-representative of~$j$, and we make all the nodes in~$B_N$ adjacent to all the dummy vertices~$d_i$ for~$i \in [t]$.
		
		\item The edge representatives~$B_E$ contain one vertex for each possible edge in an undirected~$n$-vertex graph. So for~$\{v,w\} \in \binom{[n]}{2}$ we have a vertex~$e_{v,w}$ of weight two. Vertex~$e_{v,w}$ is adjacent to an $A$-representative~$v_{i,j}$ if instance~$G_i$ contains the edge~$\{v,w\}$ and~$j = v$ or~$j = w$, i.e., the edge representative~$e_{v,w}$ is adjacent to instance $i$'s $A$-representatives of the endpoints of the edge, provided that instance~$i$ actually contains the edge. Additionally, all vertices of~$B_E$ are adjacent to all dummy vertices~$d_i$ for~$i \in [t]$.
	\end{itemize}
	The construction is completed by turning~$B := B_I \cup B_N \cup B_E$ into a clique. We set~$k' := t \cdot (n^4 + n^6) + n^3 + n^5 \log t + k - 1$.
\end{itemize}

To complete the first stage, we need to prove that~$(G^*,w^*)$ has weighted treewidth at most~$k'$ if and only if at least one of the input graphs~$G_i$ has cutwidth at most~$k$. Before proving this claim, we establish some properties of the constructed instance~$(G^*,w^*,k')$.

\begin{claim} \label{firstEliminationIsCutwidth}
Let~$S := \{ v_{i,j} \mid j \in [n] \}$ for a given instance number~$i \in [t]$ be the subset of the vertices in~$A$ corresponding to instance~$i$. Let~$\pi \colon S \to [n]$ be a permutation of~$S$. Consider the process of eliminating the vertices in~$S$ from graph~$G^*$ in the order given by~$\pi$, and let~$\eweight( \pi^{-1}(j) )$ be the total weight of~$N[\pi^{-1}(j)]$ when eliminating the vertex~$\pi^{-1}(j)$ for~$j \in [n]$. Then~$\eweight(\pi^{-1}(j)) = t \cdot (n^4 + n^6) + n^3 + n^5 \log t + \ell$, where~$\ell := | \{ \{u,v\} \in E(G_i) \mid \pi(u) \leq j < \pi(v) \}|$.
\end{claim}
\begin{proof}
The intuition behind the proof is that the elimination process has two effects on the weight of neighbors of some vertex~$v\in A$: on the one hand, eliminated vertices in~$A$ are essentially replaced by the representatives in~$B_N$ in the neighborhood of~$v$, which have slightly smaller weight than the originals; the difference is exactly equal to the degree of the corresponding vertex. On the other hand, the representative of any edge in~$B_E$ will be added to those neighborhoods, once one of the endpoints is eliminated; recall that those edges contribute a weight of two. Thus, when reaching the first endpoint of an edge, the weight increases by one (by the degree contribution); when reaching the second endpoint this increase is canceled. Together this leads to the contribution of~$\ell$ in~$\eweight(\pi^{-1}(j))$. This idea was used by Arnborg et al.~\cite{ArnborgCP87} in their NP-completeness proof for \treewidth.

Armed with this intuition, let us proceed with the proof. By definition of~$G^*$, all vertices in~$S$ have the same set of neighbors in~$B_I$ so elimination of vertices from~$S$ does not affect the adjacency of other vertices in~$S$ to~$B_I$. Consider a vertex~$v_{i,j}$ in~$S$. From the construction of~$G^*$ it follows that initially, the only vertex of~$S$ which is adjacent to the $B$-representative of~$j$, is the vertex~$v_{i,j}$. Since we only eliminate vertices from~$S$, it follows that a vertex in~$S$ is only adjacent to the~$B$-representative of a node number~$j$ if that vertex is itself the unique $A$-representative of~$j$ in~$S$, or if the $A$-representative of~$j$ in~$S$ was eliminated earlier. Let us use these observations to prove the claim.

For an arbitrary value of~$j \in [n]$ we consider the closed neighborhood of the vertex~$\pi^{-1}(j)$ just before it is eliminated. We will study the neighborhood of~$\pi^{-1}(j)$ in the sets~$A, B_I, B_N$ and~$B_E$ consecutively. For convenience, define $\eweight^X (\pi^{-1}(j) )$ for~$X \subseteq V(G^*)$ as the total weight of~$N[\pi^{-1}(j)] \cap X$ when~$\pi^{-1}(j)$ is eliminated.
\begin{description}
	\item[Neighbors in~$A$.] Since~$A$ is a clique and~$S \subseteq A$, vertex~$\pi^{-1}(j)$ is initially adjacent to all vertices of~$A$. Since the only vertices which are eliminated are those in~$S$ corresponding to instance~$i$, vertex~$\pi^{-1}(j)$ will be adjacent to all vertices for other instances, i.e., to~$v_{i', j}$ for~$i' \neq i$ and~$j \in [n]$, for a total weight of~$(t-1) n\cdot n^3$. Vertex~$\pi^{-1}(j)$ is also adjacent to all~$t$ dummy vertices for a weight of~$n^6$ each. The remaining vertices of~$A$ are those in~$S$, and~$\pi^{-1}(j)$ is adjacent to those which are not already eliminated. Hence there are~$n - j + 1$ vertices in~$S$ which are in the closed neighborhood of~$\pi^{-1}(j)$ just before it is eliminated. These have weight~$n^3 (n - j + 1)$ so~$\eweight^{A}(\pi^{-1}(j)) = (t-1) n^4 + t\cdot n^6 + n^3(n - j + 1)$.
	\item[Neighbors in~$B_I$.] Since the neighborhood of~$\pi^{-1}(j)$ in~$B_I$ is not changed by the eliminations, vertex~$\pi^{-1}(j)$ has exactly~$\log t$ neighbors in~$B_I$ with weight~$n^5$ each so~$\eweight^{B_I}(\pi^{-1}(j)) \linebreak[1] = n^5 \log t$.
	\item[Neighbors in~$B_N$.] By construction of~$G^*$, vertex~$\pi^{-1}(j)$ is initially adjacent to the unique node in~$B_N$ that is the $B$-representative of~$\pi^{-1}(j)$, and to no other vertices of~$B_N$. For each vertex~$1 \leq j' < j$ that was eliminated before~$j$, vertex~$\pi^{-1}(j)$ has become adjacent to the $B$-representative of~$\pi^{-1}(j')$ in~$B_N$. So $\eweight^{B_N}(\pi^{-1}(j)) = \sum _{j'=1}^{j} (n^3 - \deg(\pi^{-1}(j')))$.
	\item[Neighbors in~$B_E$.] Initially, vertex~$\pi^{-1}(j)$ is adjacent to the edge-representative vertices in~$B_E$ for which~$\pi^{-1}(j)$ represents an endpoint, so to~$\deg(\pi^{-1}(j))$ vertices with weight two each. For each vertex~$\pi^{-1}(j')$ with~$1 \leq j' < j$ that is eliminated before~$\pi^{-1}(j)$, vertex~$\pi^{-1}(j)$ becomes adjacent to the edge-representatives in~$B_E$ for edges that are incident on~$\pi^{-1}(j')$ in graph~$G_i$. This shows that $\eweight^{B_E}(\pi^{-1}(j)) = 2 | \bigcup _{1 \leq j' \leq j} \{ e_{u,v} \mid \{u,v\} \in E(G_i) \wedge (j' = u \vee j' = v)\} |$.
\end{description}
We can now sum up the weights of the members of the closed neighborhood $N[\pi^{-1}(j)]$ in each of the respective subsets to establish that~$\eweight(\pi^{-1}(j))$ equals:
\allowdisplaybreaks
\begin{align*}
%& \eweight(\pi^{-1}(j)) \\
& \eweight^A(\pi^{-1}(j)) + \eweight^{B_I \cup B_N \cup B_E}(\pi^{-1}(j)) \\
=& [(t-1) n^4 + t \cdot n^6 +  n^3(n - j + 1)] + [n^5 \log t] + \left [\sum _{j'=1}^{j} (n^3 - \deg(\pi^{-1}(j'))) \right] + \\ 
& 2 \left | \bigcup _{1 \leq j' \leq j} \{ e_{u,v} \mid \{u,v\} \in E(G_i) \wedge (j' = u \vee j' = v)\} \right | \\
=& t \cdot (n^4 + n^6) + n^3 + n^5 \log t - \sum _{j'=1}^{j} \deg(\pi^{-1}(j')) + \\
& 2 \left | \bigcup _{1 \leq j' \leq j} \{ e_{u,v} \mid \{u,v\} \in E(G_i) \wedge (j' = u \vee j' = v)\} \right |.
\intertext{To simplify this further, we define~$E_1$ as the set of edges of~$G_i$ that have one endpoint among the vertices represented by~$\{\pi^{-1}(1), \ldots, \pi^{-1}(j)\}$, and~$E_2$ as the edges of~$G_i$ with both endpoints represented by~$\{\pi^{-1}(1), \ldots, \pi^{-1}(j)\}$. Observe that these definitions imply that~$\sum _{j'=1}^j \deg(\pi^{-1}(j')) = |E_1| + 2|E_2|$ and~$| \bigcup _{1 \leq j' \leq j} \{ e_{u,v} \mid \{u,v\} \in E(G_i) \wedge (j' = u \vee j' = v)\} | = |E_1| + |E_2|$. We continue the derivation:}
=& t \cdot (n^4 + n^6) + n^3 + n^5 \log t - (|E_1| + 2|E_2|) + 2(|E_1| + |E_2|) \\
=& t \cdot (n^4 + n^6) + n^3 + n^5 \log t + |E_1|.
\end{align*}
Now observe that by definition, $|E_1|$ is the number of edges that have exactly one endpoint among the first~$j$ vertices to be eliminated (and whose other endpoint is therefore eliminated later). Hence~$|E_1|$ is exactly the value of~$\ell$ as defined in the statement of the claim; this concludes the proof of \claimref{firstEliminationIsCutwidth}.
\end{proof}

The preceding claim relates the cost of the first~$n$ eliminations of an ordering of~$G^*$ to the cutwidth of an instance~$i$, provided that the ordering starts by eliminating the $A$-representatives of instance~$i$. The next claim shows that these first~$n$ eliminations essentially dominate the cost of elimination orderings with this structure.

\begin{claim} \label{firstEliminationIsBottleneck}
Let~$S$,~$i$,~$\pi$, and~$\eweight$ be as defined in \claimref{firstEliminationIsCutwidth}. Consider an elimination ordering for~$G^*$ that starts by eliminating~$S$ in the order given by~$\pi$, then eliminates the dummy~$d_i$ corresponding to instance~$i$, and eliminates the remaining vertices in arbitrary order. The cost of~$\pi$ is~$\max_{j \in [n]} \eweight(\pi^{-1}(j))$.
\end{claim}
\begin{proof}
By \claimref{firstEliminationIsCutwidth}, the maximum weight of a closed neighborhood when eliminating the vertices from~$S$ is exactly~$\max_{j \in [n]} \eweight(\pi^{-1}(j)) \geq t \cdot (n^4 + n^6) + n^3 + n^5 \log t$. We show that after elimination of~$S$, eliminating the dummy~$d_i$ and all remaining vertices does not incur a cost higher than this.

Consider the weight of the closed neighborhood of the dummy vertex~$d_i$ after the~$n$ vertices from~$S$ have been eliminated. 
At that stage,~$d_i$ is adjacent to all vertices that are left in~$A$, to all vertices of~$B_N$, some vertices of~$B_E$, and to the~$\log t$ vertices in~$B_I$ which correspond to the binary representation of the number~$i$. Since the total weight of~$B_N$ is not more than~$n \cdot n^3$, the weight of~$N[d_i]$ when~$d_i$ is eliminated is bounded by~$t(n^4 + n^6) - n^4 + n \cdot n^3 + n^5 \log t + 2 \binom{n}{2}$, which is at most~$t (n^4 + n^6) + n^3 + n^5 \log t$ and so does not exceed the cost incurred for the first~$n$ vertices as~$n \geq 2$.

After~$d_i$ and~$S$ have been eliminated from the graph, the total weight of the remaining vertices is at most~$(t-1)(n^4 + n^6) + 2 n^5 \log t + n^4 + 2 \binom{n}{2}$, which is bounded by $t (n^4 + n^6) + n^3 + n^5 \log t$ as we assumed~$n \geq \log t$ at the beginning of the proof. Hence the cost of this elimination ordering~$\pi$ is dominated by the cost of eliminating the first~$n$ vertices and is therefore~$\max_{j \in [n]} \eweight(\pi^{-1}(j))$.
\end{proof}

Having shown how the cost of specific types of elimination orderings of~$G^*$ corresponds to the cutwidth of one particular input instance, we proceed to show that there is always an optimal ordering of this type.

\begin{claim} \label{claim:elimOrderStructure}
If there is an elimination ordering of~$(G^*,w^*)$ of cost at most~$k'$, then there is such an ordering that starts by eliminating all vertices in the set~$\{v_{i, j} \mid j \in [n]\}$ for some~$i \in [t]$, i.e., there is an ordering that first eliminates all $A$-representatives corresponding to \emph{one} particular input instance~$G_i$.
\end{claim}
\begin{proof}
The proof contains of two parts. We first give a canonical elimination ordering of bounded cost, and then use a replacement argument which compares the cost of this canonical ordering to an ordering which does not match the form in the statement of the lemma.

Define a canonical elimination ordering~$\pi^*$ as follows. It starts with the sequence $v_{1, 1}, v_{1, 2}, \ldots, \linebreak[1] v_{1, n}$, then eliminates dummy~$d_1$, and finally eliminates the rest of the vertices in arbitrary order. By \claimref{firstEliminationIsBottleneck} the cost incurred by ordering~$\pi^*$ is $\max_{j \in [n]} \eweight(\pi^{-1}(j))$. Observe that the cutwidth of graph~$G_1$ under any ordering does not exceed the number of its edges, which is at most~$3n/2$ since~$G_1$ has maximum degree at most three. Hence we find that the term~$\ell$ in the expression for~$\eweight$ given by \claimref{firstEliminationIsCutwidth} is bounded by~$3n/2$. Using \claimref{firstEliminationIsCutwidth} we therefore find that the cost of this canonical elimination ordering~$\pi^*$ is bounded by~$t \cdot (n^4 + n^6) + n^3 + n^5 \log t + 3n/2$.

To complete the proof, the second part will show that any elimination ordering whose form does not match that in the statement of the claim, has cost as least as much as the canonical ordering. So consider any elimination ordering~$\pi$ of~$(G^*, w^*)$ of cost at most~$k'$. Since~$B$ is a clique in~$G^*$, by \proposref{eliminatePartiteSetFirst} there is an optimal-cost elimination ordering which first eliminates all of~$A$. So assume without loss of generality that~$\pi$ first eliminates all vertices of~$A$. Since the construction of~$G^*$ guarantees that for all~$i \in [t]$, all vertices~$v_{i,j}$ of~$A$ satisfy~$N_{G^*}[v_{i,j}] \subseteq N_{G^*}[d_{i}]$, \proposref{eliminateLargerNeighborhoodLater} shows that we may assume without loss of generality that for all~$i \in [t]$, the  vertices~$v_{i,j}$ for~$j \in [n]$ are eliminated by~$\pi$ earlier than~$d_i$; hence the first~$n$ vertices eliminated by~$\pi$ are not dummy vertices. If the first~$n$ vertices correspond to the same instance then we 
are done; hence in the remainder we may assume this is not the case. Consider the first index~$1 < j \leq n$ such that all vertices~$\pi^{-1}(j')$  for~$1 \leq j' < j$ correspond to the same instance~$i$ (i.e., they are of the form~$v_{i, j''}$ for~$j'' \in [n]$) and~$\pi^{-1}(j)$ corresponds to instance~$i'$ with~$i \neq i'$. Let us consider the neighborhood of the vertex~$\pi^{-1}(j)$ when it is eliminated. 

By construction of~$G^*$, vertex~$\pi^{-1}(j)$ corresponding to instance~$i'$ is adjacent to the vertices in~$B_I$ which correspond to the binary representation of~$i'$. Since vertex~$\pi^{-1}(1)$ was eliminated before~$\pi^{-1}(j)$, and since vertices~$\pi^{-1}(1)$ and~$\pi^{-1}(j)$ are adjacent in~$G^*$ because they are both members of the clique~$A$, after elimination of~$\pi^{-1}(1)$ the vertex~$\pi^{-1}(j)$ has become adjacent to all neighbors of~$\pi^{-1}(1)$. Since~$\pi^{-1}(1)$ is adjacent to the vertices of~$B_I$ corresponding to the binary representation of~$i$, and since the binary representations of~$i$ and~$i'$ must differ in at least one position, the number of neighbors of~$\pi^{-1}(j)$ in~$B_I$ at the time it is eliminated is at least~$1 + \log t$, and they have weight~$n^5$ each. Since~$\pi^{-1}(j)$ is also adjacent to all vertices of~$A$ except the~$j-1$ vertices of weight~$n^3$ which were eliminated earlier, this shows that the weight of the closed neighborhood of~$\pi^{-1}(j)$ at the time it is eliminated is at least~$t(n^4 + n^6) - j \cdot n^3 + (
1 + \log t) n^5$. Using that~$j \leq n$ and~$n \geq 2$ (which we assumed in the beginning the proof of the theorem), it now follows that the weight of~$\pi^{-1}(j)$ at the time it is eliminated is at least as much as the cost of the canonical elimination ordering. Hence the canonical elimination ordering which we defined earlier has cost no more than~$\pi$, and since the canonical ordering starts by eliminating~$v_{1, 1}, v_{1, 2}, \ldots, v_{1, n}$ this concludes the proof of \claimref{claim:elimOrderStructure}.
\end{proof}

We are now finally ready to prove that~$(G^*, w^*)$ has weighted treewidth at most~$k'$ if and only if at least one of the input graphs~$G_i$ has cutwidth at most~$k$. First assume that~$(G^*, w^*)$ has weighted treewidth at most~$k'$. By \proposref{eliminationCostIsWeightedWidth} this implies that there is an elimination ordering~$\pi$ of~$G^*$ with cost at most~$k'+1$. By \claimref{claim:elimOrderStructure} we may assume that there is an instance number~$i^* \in [t]$ such that~$\pi$ starts by eliminating all vertices in the set~$S := \{ v_{i^*, j} \mid j \in [n] \}$. As the cost of~$\pi$ is at most~$k'+1$, the weight of the closed neighborhood of a vertex in~$S$ at the time it is eliminated does not exceed~$k'+1$. By \claimref{firstEliminationIsCutwidth} this proves that~$\max _{j \in [n]} \eweight( \pi^{-1}(j) ) \leq k'+1$. Plugging in the value for~$k'$ and the expression for~$\eweight$ obtained in the mentioned claim, and canceling terms on both sides, we find that~$\max_{j \in [n]} | \{ \{u,v\} \in E(G_{i^*} ) \mid \pi(u) \leq j < \pi(v) \}| \leq k$ which proves that~$G_{i^*}$ has cutwidth at most~$k$, when using the ordering on~$S$ induced by~$\pi$.

For the reverse direction, assume that~$G_{i^*}$ has cutwidth at most~$k$, and let~$\pi^*$ be an ordering which achieves this cutwidth. Build an elimination ordering for~$G^*$ by first eliminating the vertices of~$S := \{ v_{i^*, j} \mid j \in [n] \}$ in the order induced by~$\pi^*$, then eliminating the dummy~$d_{i^*}$, and then eliminating the remaining vertices in arbitrary order. By \claimref{firstEliminationIsBottleneck} the cost of this ordering is dominated by the cost of eliminating the vertices of~$S$, which is~$\max _{j \in [n]} \eweight( \pi^{-1}(j) )$. If ordering~$\pi^*$ achieves cutwidth at most~$k$ on~$G_{i^*}$, then evaluating the expression for~$\eweight$ given by \claimref{firstEliminationIsCutwidth} proves that the cost of~$\pi$ is at most~$k' + 1$. Using \proposref{eliminationCostIsWeightedWidth} this proves that~$(G^*, w^*)$ has weighted treewidth at most~$k'$.

To complete the cross-composition of \cutwidththree into \TWMSC, we can transform the weighted graph~$(G^*, w^*)$ into the unweighted graph~$G'$ using the transformation of \proposref{treewidthVersusWeightedWidth}. Since this transformation duplicates the closed neighborhoods of vertices, it results in a co-bipartite graph: the cliques~$A$ and~$B$ of~$G^*$ are just transformed into larger cliques in~$G'$. Let~$B'$ be the clique in~$G'$ that results from the transformation of clique~$B$ in~$G^*$. The size of~$B'$ is bounded by the maximum weight of a vertex in~$B$ (under~$w^*$) times the size of~$B$. Since both are polynomial in~$n + \log t$, this shows that the size of~$B'$ is bounded polynomially in~$n + \log t$. The set~$B'$ is a modulator to a single clique in~$G'$, as~$B'$ is one of the partite sets of the co-bipartite graph. Now consider the instance of \TWMSC which asks if~$G'$ with the modulator~$B'$ to a single clique has treewidth at most~$k'$; by the equivalence between the weighted treewidth of the original graph, and the unweighted treewidth of the result of the transformation, our constructed instance is equivalent to the OR of the input instances of \cutwidththree. The size of the modulator, which is the parameter of the \TWMSC instance, is polynomial in~$n + \log t$. This concludes the cross-composition; \thmref{theorem:twmsc:nokernel} follows by applying \thmref{crossCompositionNoKernel}.
\end{proof}

Since the pathwidth of a co-bipartite graph equals its treewidth~\cite{Mohring96} and the graph formed by the cross-composition is co-bipartite, we obtain the following corollary.

\begin{corollary}
\textsc{Pathwidth parameterized by a modulator to a single clique} does not admit a polynomial kernel unless \containment.
\end{corollary}

\noindent For completeness we remark that in the extended abstract~\cite{BodlaenderJK12a} containing \thmref{theorem:twmsc:nokernel}, we also proved that \pathwidthByVC has a kernel with~$\Oh(\ell^3)$ vertices. It is currently unknown whether \pathwidthByFVS admits a polynomial kernelization.

\subsection{A kernelization lower bound for Weighted Treewidth parameterized by Vertex Cover}
\label{section:weighted} \label{sect:tw:weightedLowerBound}

The \weightedtreewidth problem asks for a given vertex-weighted graph~$G$ and integer~$k$, if the weighted treewidth of~$G$ is at most~$k$. Preprocessing heuristics for the problem were considered by Van den Eijkhof et al.~\cite{EijkhofBK07}. We consider a parameterized version of this problem, where we choose the size of a vertex cover as the parameter:

\parproblemdef{Weighted Treewidth parameterized by a vertex cover}
{A graph~$G$, a weight function~$w \colon V(G) \to \mathbb{N}$ encoded in unary, an integer~$k$, and a vertex cover~$S\subseteq V(G)$ of~$G$.}
{$\ell:=|S|$.}
{Does~$G$ have a tree decomposition of weighted width at most~$k$ with respect to~$w$?}

\noindent We show that \weightedTreewidthByVC does not admit a polynomial kernelization unless \containment and the polynomial hierarchy collapses. Our lower bound proof is based on the effect of the join operation on the treewidth. We therefore use the following lemma due to Bodlaender and M\"ohring~\cite{BodlaenderM93}.

\begin{lemma}[\cite{BodlaenderM93}]\label{lemma:treewidth:join}For any two graphs~$G_1$ and~$G_2$ it holds that:
\[
\tw(G_1\otimes G_2)=\min(\tw(G_1)+|V(G_2)|,\tw(G_2)+|V(G_1)|).
\]
\end{lemma}

\noindent From \lemmaref{lemma:treewidth:join}, we directly obtain the following corollary.

\begin{corollary}\label{corollary:graphjoin}
For any~$t$ graphs~$G_1,\ldots,G_t$ on~$n$ vertices each it holds that:
\[
\tw(G_1\otimes\ldots\otimes G_t)=(t-1)\cdot n+\min_{i\in[t]}\tw(G_i).
\]
\end{corollary}

We also use the following weighted variant of a lemma by Bodlaender~\cite{Bodlaender96}.

\begin{lemma}\label{lemma:sharedneighbors}
Let~$G$ be a graph, let~$k$ be an integer, and let~$u,v$ be two vertices of~$G$ whose shared neighbors have combined weight at least~$k+1$. Then in any tree decomposition of~$G$ of weighted width at most~$k$ there is a bag containing both~$u$ and~$v$.
\end{lemma}

\noindent As the weighted treewidth of a graph can be defined as one less than the maximum weight of a clique in a chordal supergraph~\cite[Theorem 8]{BodlaenderR03}, the correctness of the lemma follows analogously to the correctness of \ruleref{rule:neighborimprovementS}.

\begin{theorem}\label{theorem:weightedtreewidth:lowerbound}
\weightedTreewidthByVC does not admit a polynomial kernelization unless \containment.
\end{theorem}

\begin{proof}
We prove the theorem by giving a cross-composition from the classic (unweighted) \treewidth problem into \weightedTreewidthByVC.

%We define a simple polynomial equivalence relation~$\R$ such that two well-formed instances~$(G,k)$ and~$(G',k')$ of \treewidth are equivalent if the graphs~$G$ and~$G'$ have the same numbers~$n$ and~$m$ of vertices and edges, respectively, and if~$k=k'\leq n$. For technical reasons, all instances where~$k$ exceeds the number of vertices also form one equivalence class, but we will tacitly ignore these trivial instances in the following (it is easy to see that the presence of such an instance makes a cross-composition trivial). Similarly we make all strings equivalent that do not encode valid instances.  Clearly, equivalence under~\R can be checked in polynomial time and a set of well-formed instances each on at most~$n$ vertices is partitioned into at most~$n^4$ equivalence classes. As the strings that do not encode a valid instance add one class, as do the instances whose answer is trivially \yes, this yields a polynomial equivalence relation.

By a suitable choice of polynomial equivalence relation~$\R$, similar to that of \thmref{theorem:twmsc:nokernel}, it suffices to show how to compose a series of~$t$ instances~$(G_1, k), \ldots, (G_t, k)$ of the classical \treewidth problem that all ask for the same target value~$k$, and in which each input graph has the same number of~$n$ vertices and~$m$ edges. Additionally, we may assume that~$t,n \geq 10$.

%Accordingly, let~$(G_1, k_1),\ldots,(G_t,k_t)$ be~$t$ instances of \treewidth, each instance~$(G_i,k_i)$ asking whether the graph~$G_i$ on~$n$ vertices and~$m$ edges has a tree decomposition of (unweighted) width at most~$k_i = k$. We may assume that~$t,n \geq 10$, as otherwise we either solve the instances (outputting a constant-size \yes or \no-instance with the same answer) or copy one input instance a constant number of times (resulting in only a polynomial increase in~$t$ and the total input size). Hence in the remainder we assume~$t,n \geq 10$.

The cross-composition can be interpreted in terms of the graph~$H := G_1 \otimes \ldots \otimes G_t$ obtained by joining all input graphs together. Let~$k' := (t-1)n + k$. By \corollaryref{corollary:graphjoin} there is an input graph of treewidth at most~$k$ if and only if~$\tw(H) \leq k'$. The joined graph~$H$ therefore encodes the logical OR of the input instances. We want to exploit this behavior, but the graph~$H$ does not have a small vertex cover. We therefore use a construction that simulates the edges in the graph~$H$ by adding a small number of high-weight vertices to the graph, effectively allowing these edges to be dropped from~$H$ at the expense of introducing~$\Oh(n^2 \log t)$ new vertices that will form a vertex cover. Rather than replacing each edge by one high-weight vertex --- which would cause the vertex cover to become too large --- we introduce a constant number of vertices for each clique in an \emph{edge clique cover} of~$H$, and subsequently show that~$H$ has a small edge clique cover.
 Recall that an edge clique cover of a graph is a collection of cliques such that for each edge of the graph, there is a clique in the cover that contains both its endpoints.

\begin{claim} \label{claim:replaceECCC}
Let~$(G,k)$ be an instance of \treewidth, and let~$\C = \{C_1, \ldots, C_r\}$ be an edge clique cover of~$G$ whose largest clique has size at most~$c$. Let~$d$ be a positive integer, and let~$G'$ be the graph with weight function~$w'$ constructed as follows:
\begin{itemize}
	\item Initialize~$G'$ as a graph without edges on the vertex set~$V(G)$. Set the weight of all vertices to one.
	\item For each~$i \in [r]$ add two vertices to~$G'$, of weight~$d$, with open neighborhood~$C_i$.
\end{itemize}
If~$2d > k$ and~$c+d \leq k$, then the weighted treewidth of~$G'$ is at most~$k$ if and only if~$\tw(G) \leq k$. Moreover, the~$2r$ newly added vertices~$X'$ form a vertex cover of the graph~$G'$.
\end{claim}
\begin{proof}
As all edges from~$G$ are dropped, the~$2r$ new vertices form a vertex cover of~$G'$. We prove the correspondence between the treewidth values.

($\Leftarrow$) Let~$\T = (T, \{\X_i \mid i \in V(T)\})$ be a tree decomposition of~$G$ of width~$k$. We show how to incorporate the newly added vertices in this decomposition without increasing its width. So consider a vertex~$v_i$ with open neighborhood~$C_i$, that was added to~$G'$ on behalf of a clique~$C_i$ in the cover of~$G$. As~$C_i$ is a clique in~$G$, there is a bag~$\X_i$ of~$\T$ that contains all vertices of~$C_i$ by \proposref{proposition:simpleCliqueContainment}. Hence we may accommodate~$v_i$ by creating a new bag containing~$C_i \cup \{v_i\}$, and making the bag adjacent to~$\X_i$. Let~$\T'$ be the result of consecutively adding all new vertices to~$\T$ in this way. As each of the newly added vertices occurs in only one bag, we satisfy the convexity property and obtain a valid tree decomposition of~$G'$. Each new bag that we introduce has weight at most~$c + d \leq k$ since each clique has at most~$c$ vertices, all of weight one, and newly added vertices have weight~$d$. Therefore the decomposition has weighted width no larger than~$k$. We conclude that the weighted treewidth of~$G'$ is at most~$k$.

($\Rightarrow$) Consider a tree decomposition~$\T' = (T, \{\X_i \mid i \in V(T)\})$ of~$G'$ having weighted width at most~$k$. We show that restricting each bag to the vertex set of~$G$ yields a valid tree decomposition of~$G$. As the weighted width of~$\T'$ is at most~$k$, and each vertex has weight at least one, this restriction has unweighted width at most~$k$. It is easy to see that the restriction operation satisfies the convexity properties of a tree decomposition, and that each vertex of~$G$ is present in some bag. It therefore suffices to show that for each edge~$\{u,v\}$ of~$G$, the restriction has a bag containing both endpoints. To this end, let~$C_i$ be a clique in the edge clique cover containing~$u$ and~$v$. The two vertices that were added to~$G'$ on behalf of clique~$C_i$ have combined weight~$2w$, and are adjacent to both~$u$ and~$v$. Therefore the common neighbors of~$u$ and~$v$ have weight at least~$2d > k$, which shows by \lemmaref{lemma:sharedneighbors} that~$\T'$ has a bag 
containing~$u$ and~$v$. The restriction therefore also has such a bag, which shows that all edges are covered by the decomposition.
\end{proof}

\indent The following claim shows that a small edge clique cover of~$H$ can be constructed efficiently.

\begin{claim} \label{claim:findECCC}
Given graphs~$G_1, \ldots, G_t$ on~$n$ vertices and~$m$ edges each, we can construct an edge clique cover~$\C$ of the graph~$G_1\otimes\ldots\otimes G_t$ consisting of~$\Oh(n^2 \log t)$ cliques in polynomial time, such that each clique in the cover has at most~$2t$ vertices.
\end{claim}
\begin{proof}
We first construct~$m$ cliques to cover all the edges contained in one of the input graphs; later we show how to cover the join edges that are added between the different graphs.

We number the edges within each input graph from one to~$m$, and cover them one edge index at a time. For an edge index~$j \in [m]$ let~$C_i$ contain the endpoints of the $j$-th edge in each input graph. Since vertices originating from different input graphs are made adjacent by the join, each~$C_i$ is a clique in the joined graph. Every clique created in this way has at most two vertices from each input graph, hence at most~$2t$ vertices in total. It is easy to see that all edges of the input graphs are covered by this series of~$m \leq n^2$ cliques.

To cover the join edges, we use a strategy based on binary expansions. For each~$r \in [ \lceil \log t \rceil ]$ and each combination of~$p,q \in [n]$ define a clique~$C_{r,p,q}$ as follows. The clique contains exactly one vertex from each input graph~$G_i$. If the~$r$-th bit of the binary expansion of the number~$i$ is a zero, then~$C_{r,p,q}$ contains the $p$-th vertex from~$G_i$. If the bit is a one, the clique contains the~$q$-th vertex from~$G_i$. By the nature of the join operation, all sets defined in this way are cliques. The resulting family trivially consists of~$\Oh(n^2 \log t)$ cliques. It remains to show that all edges between different input graphs are covered by the family.

To see this, consider an edge between the~$p$-th vertex of input~$G_i$, and the $q$-th vertex of input~$G_{i'}$. As~$i \neq i'$ the binary expansions of~$i$ and~$i'$ differ in at least one bit position~$r$. From the given construction it now follows easily that either~$C_{r,p,q}$ or~$C_{r,q,p}$ contains both vertices. Hence for each join edge there is a clique that covers them. As it is easy to see that the construction can be performed in polynomial time, this concludes the proof.
\end{proof}

The final construction combines these two claims. Presented with the inputs~$(G_1, k), \ldots, (G_t, k)$ the cross-composition builds the graph~$H = G_1 \otimes \ldots \otimes G_t$, and uses \claimref{claim:findECCC} to find an edge clique cover~$\C$ of~$H$ consisting of~$r \in \Oh(n^2 \log t)$ cliques, each of size at most~$c := 2t$. Recall that~$k' = (t-1)n + k$, and pick~$d := k' - 2t$. Using~$t,n \geq 10$ it is straight-forward to verify that~$d \geq 1$ and~$2d > k'$. It is trivial that~$c+d = k'$, which shows that we may apply \claimref{claim:replaceECCC} to~$G_1, \ldots, G_t$ and~$\C$ with this choice of~$d$. Applying the construction in the claim, the algorithm obtains a weighted graph~$(G',w')$ with a vertex cover~$S'$ of size~$2r \in \Oh(n^2 \log t)$, which has weighted treewidth at most~$k'$ if and only if~$H$ has treewidth at most~$k'$. By \lemmaref{lemma:treewidth:join} the latter happens if and only if~$\tw(G_i) \leq k$ for some~$i \in [t]$. Hence the instance~$(G', S', k', w')$ is a valid 
output instance of \weightedTreewidthByVC with~$|S'| \in \Oh(n^2 \log t)$. As the construction can be carried out in polynomial time, and the output parameter is suitably bounded, this construction implies the theorem using \thmref{crossCompositionNoKernel}.
\end{proof}

\subsection{FPT classifications for (Weighted) Treewidth with structural parameterizations} \label{sect:tw:fptClassifications}
We justify the kernelization lower bounds given in this section by showing that the corresponding problems are fixed-parameter tractable. Towards this end we provide a short argument based on modular decomposition (cf.~\cite[Section 1.5]{BrandstadtLS99}). Recall that a nontrivial module in a graph~$G$ is a nonempty vertex set~$S \subsetneq V(G)$ such that for each vertex~$u \in V(G) \setminus S$ it either holds that~$N_G(u) \cap S = S$ or~$N_G(u) \cap S = \emptyset$. A graph is \emph{prime} if it has no nontrivial modules. 
% Our classification is based on the following result due to Bodlaender and Rotics.
Our classification is based on work by Bodlaender and Rotics.

%\begin{theorem}[Compare {\cite[Corollary 13]{BodlaenderR03}}] \label{theorem:treewidthByPrimeSubgraphs}
%Let~$f \colon \mathbb{N} \times \mathbb{N} \to \mathbb{N}$ be a function such that~$f(n_1 + n_2, m_1 + m_2) \geq f(n_1, m_1) + f(n_2, m_2)$ for all natural numbers as arguments. If \textsc{Weighted Independent Treewidth} can be solved in~$f(n,m)$ time on graphs in~$\F$, then \treewidth can be solved in~$\Oh(f(n,m) + n + m)$ time on graphs whose induced prime subgraphs belong to~$\C$.
%\end{theorem}

\begin{proposition}[Compare {\cite[Corollary 13]{BodlaenderR03}}] \label{proposition:treewidthByPrimeSubgraphs}
\treewidth on an $n$-vertex graph~$G$ can be solved in polynomial time using an oracle for the \witreewidth problem on induced prime subgraphs of~$G$ whose vertices have weight at most~$n$ each.
%\treewidth on an $n$-vertex graph~$G$ can be solved in~$\Oh(f(\ell)n^{\Oh(1)})$ time if the \witreewidth problem can be solved in~$f(\ell)n^{\Oh(1)}$ time on induced $\ell$-vertex prime subgraphs of~$G$ with vertex weights at most~$n$.
%
%If \witreewidth on an $\ell$-vertex graph belonging to class~$\C$ can be solved in~$f(\ell)w^{\Oh(1)}$ time when each vertex has weight at most~$w$, then \treewidth can be solved in~$
%
%If \witreewidth on an $\ell$-vertex graph belonging to class~$\C$ can be solved in~$f(n)w^{\Oh(1)}$ time when each vertex weight it at most~$w$, then \treewidth can be solved in~$\Oh(f(n)n^{\Oh(1)})$ time on graphs whose induced prime subgraphs belong to~$\C$.
%If \textsc{Weighted Independent Treewidth} can be solved in~$f(n,m)$ time on graphs in~$\F$, then \treewidth can be solved in~$\Oh(f(n,m) + n + m)$ time on graphs whose induced prime subgraphs belong to~$\C$.
\end{proposition}

The statement is a variant of Corollary 13 in the work of Bodlaender and Rotics~\cite{BodlaenderR03}, tailored towards FPT classifications. The precise definition of \witreewidth is not important for our purposes; it suffices to know that it is a problem on weighted graphs, that can be solved by doing a polynomial-time weight computation for each chordal supergraph of the input. Hence there is a computable function~$f$ such that \witreewidth can be solved in~$f(\ell)w^{\Oh(1)}$ time on graphs with~$\ell$ vertices, each having weight at most~$w$. To leverage \proposref{proposition:treewidthByPrimeSubgraphs} to obtain FPT algorithms for structural parameterizations of \treewidth, the following result relates the sizes of prime subgraphs and vertex covers.

\begin{proposition} \label{proposition:primeSubgraphsByVC}
If~$S \subseteq V(G)$ is a vertex cover of~$G$, then no prime subgraph of~$G$ has more than~$|S| + 2^{|S|}$ vertices.
\end{proposition}
\begin{proof}
If~$H$ is a prime subgraph of~$G$ with more than~$|S| + 2^{|S|}$ vertices, then~$H$ contains at least~$2^{|S|}+1$ vertices from~$V(G) \setminus S$. Since the number of possible neighborhoods of~$V(G) \setminus S$ into~$S$ is bounded by~$2^{|S|}$, such a subgraph~$H$ contains two vertices~$u,v \not \in S$ with the same neighborhood into~$S$. As~$u$ and~$v$ are not contained in the vertex cover, all their neighbors belong to~$S$. It follows that~$\{u,v\}$ is a nontrivial module of~$H$, contradicting the assumption that it is prime.
\end{proof}

\begin{corollary}
The problems \treewidthByClique and \weightedTreewidthByVC are fixed-parameter tractable.
\end{corollary}
\begin{proof}
We combine the two propositions, using the existence of an algorithm that solves \witreewidth on $\ell$-vertex graphs with weights at most~$w$ in~$f(\ell)w^{\Oh(1)}$ time.

Observe that a graph is prime if and only if its edge-complement is prime. Since a modulator to a clique~$S$ in~$G$ forms a vertex cover in~$\overline{G}$, \proposref{proposition:primeSubgraphsByVC} bounds the size of a prime subgraph of~$G$ in terms of its deletion distance to a clique. For an instance~$(G,k,S)$ of \treewidthByClique we therefore know that all induced prime subgraphs of~$G$ have at most~$|S| + 2^{|S|}$ vertices. Hence an oracle for computing \witreewidth on induced prime subgraphs of~$G$ whose weights are at most~$n$, can be implemented in~$f(|S| + 2^{|S|})n^{\Oh(1)}$ time. Applying \proposref{proposition:treewidthByPrimeSubgraphs} therefore classifies \treewidthByClique as FPT.

When it comes to the weighted variant of treewidth, observe that by \proposref{treewidthVersusWeightedWidth} the weighted treewidth of~$G$ with weight function~$w$ equals the (unweighted) treewidth of the graph~$G'$ obtained by replacing each vertex~$v \in V(G)$ by a clique on~$w(v)$ vertices. Such a replacement step does not affect the size of prime subgraphs: no induced prime subgraph can contain two members of the same replacement clique, as they would form a nontrivial module. \proposref{proposition:primeSubgraphsByVC} therefore shows that if the weighted graph~$G$ has a vertex cover~$S$, then the size of induced prime subgraphs of the corresponding unweighted graph~$G'$ is at most~$|S| + 2^{|S|}$. By the definition of \weightedtreewidth the instances of that problem encode the vertex weights in unary, implying that~$G'$ can be obtained efficiently and that~$|V(G')|$ is polynomial in the size of~$(G,w)$. We therefore obtain an FPT-algorithm for \weightedTreewidthByVC by invoking \proposref{proposition:treewidthByPrimeSubgraphs} on~$G'$, using the fact that we can give an FPT-time oracle for \witreewidth on the induced prime subgraphs of~$G'$ as they are guaranteed to be small.
\end{proof}

The approach used here cannot establish fixed-parameter tractability for \treewidthByCograph. We can show that there are prime graphs~$G$ of arbitrary size, that can be turned into cographs by just two vertex deletions. Hence there is no analogue of  \proposref{proposition:primeSubgraphsByVC} in the cograph setting.

\section{Conclusions}
\label{section:conclusions}
We considered different parameterizations for the \treewidth problem and obtained both positive and negative results for the existence of polynomial kernels. Our first positive result, a cubic-vertex kernel for \treewidthByVC, is interesting as its algorithm largely consists of elements of existing preprocessing heuristics for \treewidth, and thus also sheds some light on the experimentally observed success of these heuristics. 

Our second positive result, a polynomial kernel for \treewidthByFVS, is not only interesting from a theoretical point of view but also has practical potential. We expect that some of the reduction rules --- in particular, our new rule for almost simplicial vertices of degree~$k+1$ (\ruleref{rule:kp1almostsimplicial}) --- are effective when preprocessing real-world inputs. Initial experiments are encouraging, but a detailed experimental study remains to be done. This could be compared to the experimental studies conducted earlier~\cite{BodlaenderKE05,EijkhofBK07}.

The lower bound for the parameterization by deletion distance from a clique rules out polynomial kernels for \treewidth for a number of possibly interesting parameters; among them are the distance from cographs, cluster, co-cluster, chordal, interval, and split graphs. Furthermore, our lower bound for \weightedTreewidthByVC rules out essentially all hope for useful kernelization results for the \weightedtreewidth problem, since neither almost-complete nor almost-independent graphs appear to admit a polynomial kernelization.

Apart from improving the kernel sizes (e.g., for parameterization by a feedback vertex set), or giving polynomial lower bounds (e.g.,~$\Omega(\ell^2)$), it seems interesting whether parameter-wise one can obtain stronger results. The hierarchy of \imgref{figure:treewidthHierarchyPicture} points to a number of open problems in this direction. For example, is there a polynomial kernel with respect to the size of a modulator to outerplanar graphs? Several other problems such as \oddcycletransversal~\cite{JansenK12}, \HamCycle~\cite{BodlaenderJK12c}, and \ThreeColoring~\cite{JansenK11b}, do not admit polynomial kernels for this parameter unless \containment. To gain a better understanding of the influence of graph structure on the difficulty of \treewidth instances, it would also be interesting to obtain complexity classifications for the parameterizations by distance from chordal graphs or cographs.

% According to Hans, people at ICALP are likely not to care about probabilistic networks; but it might be nice to put in a full version.
%Another related question is the following. Consider the {\sc Probabilistic Inference} problem for probabilistic networks. Two known algorithms for this problem are Pearl's algorithm, that uses time, exponential in the size of a \emph{loop cutset} \cite{Pearl88}, but linear in the number of variables, and the algorithm by Lauritzen and Spiegelhalter \cite{LauritzenS88}, that uses time, exponential in the treewidth of the moralization of the network. As loop cutsets are strongly related to feedback vertex sets (and see also \cite{BodlaendervD10} where kernels for these problems are related), can we also obtain results relevant for the area of probabilistic networks here?

% \bibliography{../definitions,../papers,../extra}%,../Paper}
\bibliography{def,treewidth}

\begin{thebibliography}{10}

\bibitem{ArnborgCP87}
S.~Arnborg, D.~G. Corneil, and A.~Proskurowski.
\newblock Complexity of finding embeddings in a {$k$}-tree.
\newblock {\em SIAM J. Algebra. Discr.}, 8(2):277--284, 1987.
\newblock \href {http://dx.doi.org/10.1137/0608024}
  {\path{doi:10.1137/0608024}}.

\bibitem{BeckerG96a}
A.~Becker and D.~Geiger.
\newblock Optimization of {P}earl's method of conditioning and greedy-like
  approximation algorithms for the vertex feedback set problem.
\newblock {\em Artif. Intell.}, 83(1):167--188, 1996.
\newblock \href {http://dx.doi.org/10.1016/0004-3702(95)00004-6}
  {\path{doi:10.1016/0004-3702(95)00004-6}}.

\bibitem{Bodlaender96}
H.~L. Bodlaender.
\newblock A linear-time algorithm for finding tree-decom\-positions of small
  treewidth.
\newblock {\em SIAM J. Comput.}, 25(6):1305--1317, 1996.
\newblock \href {http://dx.doi.org/10.1145/167088.167161}
  {\path{doi:10.1145/167088.167161}}.

\bibitem{Bodlaender98}
H.~L. Bodlaender.
\newblock A partial {$k$}-arboretum of graphs with bounded treewidth.
\newblock {\em Theor. Comput. Sci.}, 209(1-2):1--45, 1998.
\newblock \href {http://dx.doi.org/10.1016/S0304-3975(97)00228-4}
  {\path{doi:10.1016/S0304-3975(97)00228-4}}.

\bibitem{Bodlaender2003}
H.~L. Bodlaender.
\newblock Necessary edges in $k$-chordalizations of graphs.
\newblock {\em J. Comb. Optim.}, 7(3):283--290, 2003.
\newblock \href {http://dx.doi.org/10.1023/A:1027320705349}
  {\path{doi:10.1023/A:1027320705349}}.

\bibitem{Bodlaender09}
H.~L. Bodlaender.
\newblock Kernelization: New upper and lower bound techniques.
\newblock In {\em Proc. IWPEC 2009}, volume 5917 of {\em LNCS}, pages 17--37,
  2009.
\newblock \href {http://dx.doi.org/10.1007/978-3-642-11269-0_2}
  {\path{doi:10.1007/978-3-642-11269-0_2}}.

\bibitem{BodlaenderDFH09}
H.~L. Bodlaender, R.~G. Downey, M.~R. Fellows, and D.~Hermelin.
\newblock On problems without polynomial kernels.
\newblock {\em J. Comput. Syst. Sci.}, 75(8):423--434, 2009.
\newblock \href {http://dx.doi.org/10.1016/j.jcss.2009.04.001}
  {\path{doi:10.1016/j.jcss.2009.04.001}}.

\bibitem{BodlaenderFKKT12}
H.~L. Bodlaender, F.~V. Fomin, A.~M. C.~A. Koster, D.~Kratsch, and D.~M.
  Thilikos.
\newblock A note on exact algorithms for vertex ordering problems on graphs.
\newblock {\em Theory Comput. Syst.}, 50(3):420--432, 2012.
\newblock \href {http://dx.doi.org/10.1007/s00224-011-9312-0}
  {\path{doi:10.1007/s00224-011-9312-0}}.

\bibitem{BodlaenderFKKT12a}
H.~L. Bodlaender, F.~V. Fomin, A.~M. C.~A. Koster, D.~Kratsch, and D.~M.
  Thilikos.
\newblock On exact algorithms for treewidth.
\newblock {\em ACM Trans. Algorithms}, 9(1):12, 2012.
\newblock \href {http://dx.doi.org/10.1145/2390176.2390188}
  {\path{doi:10.1145/2390176.2390188}}.

\bibitem{BodlaenderJK11}
H.~L. Bodlaender, B.~M.~P. Jansen, and S.~Kratsch.
\newblock Cross-composition: A new technique for kernelization lower bounds.
\newblock In {\em Proc. STACS 2011}, volume~9 of {\em LIPIcs}, pages 165--176,
  2011.
\newblock \href {http://dx.doi.org/10.4230/LIPIcs.STACS.2011.165}
  {\path{doi:10.4230/LIPIcs.STACS.2011.165}}.

\bibitem{BodlaenderJK11b}
H.~L. Bodlaender, B.~M.~P. Jansen, and S.~Kratsch.
\newblock Preprocessing for treewidth: A combinatorial analysis through
  kernelization.
\newblock In {\em Proc. ICALP 2011}, volume 6755 of {\em LNCS}, pages 437--448,
  2011.
\newblock \href {http://dx.doi.org/10.1007/978-3-642-22006-7_37}
  {\path{doi:10.1007/978-3-642-22006-7_37}}.

\bibitem{BodlaenderJK11a}
H.~L. Bodlaender, B.~M.~P. Jansen, and S.~Kratsch.
\newblock Preprocessing for treewidth: A combinatorial analysis through
  kernelization.
\newblock {\em CoRR}, abs/1104.4217v1, 2011.
\newblock \href {http://arxiv.org/abs/1104.4217v1} {\path{arXiv:1104.4217v1}}.

\bibitem{BodlaenderJK12c}
H.~L. Bodlaender, B.~M.~P. Jansen, and S.~Kratsch.
\newblock Kernel bounds for path and cycle problems.
\newblock {\em Theor. Comput. Sci.}, 2012.
\newblock Online First.
\newblock \href {http://dx.doi.org/10.1016/j.tcs.2012.09.006}
  {\path{doi:10.1016/j.tcs.2012.09.006}}.

\bibitem{BodlaenderJK12a}
H.~L. Bodlaender, B.~M.~P. Jansen, and S.~Kratsch.
\newblock Kernel bounds for structural parameterizations of pathwidth.
\newblock In {\em Proc. SWAT 2012}, volume 7357 of {\em LNCS}, pages 352--363,
  2012.
\newblock \href {http://dx.doi.org/10.1007/978-3-642-31155-0_31}
  {\path{doi:10.1007/978-3-642-31155-0_31}}.

\bibitem{BodlaenderK06c}
H.~L. Bodlaender and A.~M. C.~A. Koster.
\newblock Safe separators for treewidth.
\newblock {\em Discrete Math.}, 306(3):337--350, 2006.
\newblock \href {http://dx.doi.org/10.1016/j.disc.2005.12.017}
  {\path{doi:10.1016/j.disc.2005.12.017}}.

\bibitem{BodlaenderKE05}
H.~L. Bodlaender, A.~M. C.~A. Koster, and F.~v.~d. Eijkhof.
\newblock Preprocessing rules for triangulation of probabilistic networks.
\newblock {\em Comput. Intell.}, 21(3):286--305, 2005.
\newblock \href {http://dx.doi.org/10.1111/j.1467-8640.2005.00274.x}
  {\path{doi:10.1111/j.1467-8640.2005.00274.x}}.

\bibitem{BodlaenderKW05}
H.~L. Bodlaender, A.~M. C.~A. Koster, and T.~Wolle.
\newblock Contraction and treewidth lower bounds.
\newblock {\em J. Graph Algorithms Appl.}, 10(1):5--49, 2006.
\newblock \href {http://dx.doi.org/10.7155/jgaa.00117}
  {\path{doi:10.7155/jgaa.00117}}.

\bibitem{BodlaenderM93}
H.~L. Bodlaender and R.~H. M{\"{o}}hring.
\newblock The pathwidth and treewidth of cographs.
\newblock {\em SIAM J. Discrete Math.}, 6(2):181--188, 1993.
\newblock \href {http://dx.doi.org/10.1137/0406014}
  {\path{doi:10.1137/0406014}}.

\bibitem{BodlaenderR03}
H.~L. Bodlaender and U.~Rotics.
\newblock Computing the treewidth and the minimum fill-in with the modular
  decomposition.
\newblock {\em Algorithmica}, 36(4):375--408, 2003.
\newblock \href {http://dx.doi.org/10.1007/s00453-003-1026-5}
  {\path{doi:10.1007/s00453-003-1026-5}}.

\bibitem{BrandstadtLS99}
A.~Brandst\"{a}dt, V.~Le, and J.~P. Spinrad.
\newblock {\em Graph Classes. {A} Survey}.
\newblock SIAM Monographs on Discrete Mathematics and Applications. SIAM, 1999.

\bibitem{ClautiauxCMN03}
F.~Clautiaux, J.~Carlier, A.~Moukrim, and S.~N{\`e}gre.
\newblock New lower and upper bounds for graph treewidth.
\newblock In {\em Proc. WEA 2003}, volume 2647 of {\em LNCS}, pages 70--80,
  2003.
\newblock \href {http://dx.doi.org/10.1007/3-540-44867-5_6}
  {\path{doi:10.1007/3-540-44867-5_6}}.

\bibitem{ClautiauxMNC04}
F.~Clautiaux, A.~Moukrim, S.~N{\'e}gre, and J.~Carlier.
\newblock Heuristic and meta-heuristic methods for computing graph treewidth.
\newblock {\em RAIRO Rech. Op\'{e}r.}, 38(1):13--26, 2004.
\newblock \href {http://dx.doi.org/10.1051/ro:2004011}
  {\path{doi:10.1051/ro:2004011}}.

\bibitem{Diestel10}
R.~Diestel.
\newblock {\em Graph Theory}.
\newblock Springer-Verlag, Heidelberg, 4th edition, 2010.

\bibitem{Dirac61}
G.~A. Dirac.
\newblock On rigid circuit graphs.
\newblock {\em Abh. Math. Sem. Hamburg}, 25(1-2):71--76, 1961.
\newblock \href {http://dx.doi.org/10.1007/BF02992776}
  {\path{doi:10.1007/BF02992776}}.

\bibitem{DowneyF99}
R.~Downey and M.~R. Fellows.
\newblock {\em Parameterized Complexity}.
\newblock Monographs in Computer Science. Springer, New York, 1999.

\bibitem{Drucker12_focs}
A.~Drucker.
\newblock New limits to classical and quantum instance compression.
\newblock In {\em Proc. FOCS 2012}, pages 609--618, 2012.
\newblock \href {http://dx.doi.org/10.1109/FOCS.2012.71}
  {\path{doi:10.1109/FOCS.2012.71}}.

\bibitem{EijkhofBK07}
F.~v.~d. Eijkhof, H.~L. Bodlaender, and A.~M. C.~A. Koster.
\newblock Safe reduction rules for weighted treewidth.
\newblock {\em Algorithmica}, 47(2):138--158, 2007.
\newblock \href {http://dx.doi.org/10.1007/s00453-006-1226-x}
  {\path{doi:10.1007/s00453-006-1226-x}}.

\bibitem{FellowsJR12}
M.~R. Fellows, B.~M.~P. Jansen, and F.~A. Rosamond.
\newblock Towards fully multivariate algorithmics: {Parameter} ecology and the
  deconstruction of computational complexity.
\newblock {\em European J. Combin.}, 34(3):541 -- 566, 2013.
\newblock \href {http://dx.doi.org/10.1016/j.ejc.2012.04.008}
  {\path{doi:10.1016/j.ejc.2012.04.008}}.

\bibitem{FlumG06}
J.~Flum and M.~Grohe.
\newblock {\em Parameterized Complexity Theory}.
\newblock Springer-Verlag New York, Inc., 2006.

\bibitem{FortnowS11}
L.~Fortnow and R.~Santhanam.
\newblock Infeasibility of instance compression and succinct {PCP}s for {NP}.
\newblock {\em J. Comput. Syst. Sci.}, 77(1):91--106, 2011.
\newblock \href {http://dx.doi.org/10.1016/j.jcss.2010.06.007}
  {\path{doi:10.1016/j.jcss.2010.06.007}}.

\bibitem{GuoN07}
J.~Guo and R.~Niedermeier.
\newblock Invitation to data reduction and problem kernelization.
\newblock {\em SIGACT News}, 38(1):31--45, 2007.
\newblock \href {http://dx.doi.org/10.1145/1233481.1233493}
  {\path{doi:10.1145/1233481.1233493}}.

\bibitem{JansenK11b}
B.~M.~P. Jansen and S.~Kratsch.
\newblock Data reduction for graph coloring problems.
\newblock In {\em Proc. FCT 2011}, volume 6914 of {\em LNCS}, pages 90--101,
  2011.
\newblock \href {http://dx.doi.org/10.1007/978-3-642-22953-4_8}
  {\path{doi:10.1007/978-3-642-22953-4_8}}.

\bibitem{JansenK12}
B.~M.~P. Jansen and S.~Kratsch.
\newblock On polynomial kernels for structural parameterizations of odd cycle
  transversal.
\newblock In {\em Proc. IPEC 2011}, volume 7112 of {\em LNCS}, pages 132--144,
  2011.
\newblock \href {http://dx.doi.org/10.1007/978-3-642-28050-4_11}
  {\path{doi:10.1007/978-3-642-28050-4_11}}.

\bibitem{KleitmanW91}
D.~J. Kleitman and D.~B. West.
\newblock Spanning trees with many leaves.
\newblock {\em SIAM J. Discret. Math.}, 4(1):99--106, 1991.
\newblock \href {http://dx.doi.org/10.1137/0404010}
  {\path{doi:10.1137/0404010}}.

\bibitem{KloksK95}
T.~Kloks and D.~Kratsch.
\newblock Treewidth of chordal bipartite graphs.
\newblock {\em J. Algorithms}, 19(2):266--281, 1995.
\newblock \href {http://dx.doi.org/10.1006/jagm.1995.1037}
  {\path{doi:10.1006/jagm.1995.1037}}.

\bibitem{LagergrenA91}
J.~Lagergren and S.~Arnborg.
\newblock Finding minimal forbidden minors using a finite congruence.
\newblock In {\em Proc. ICALP 1991}, volume 510 of {\em LNCS}, pages 532--543,
  1991.
\newblock \href {http://dx.doi.org/10.1007/3-540-54233-7_161}
  {\path{doi:10.1007/3-540-54233-7_161}}.

\bibitem{Mohring96}
R.~H. M{\"o}hring.
\newblock Triangulating graphs without asteroidal triples.
\newblock {\em Discrete Appl. Math.}, 64(3):281--287, 1996.
\newblock \href {http://dx.doi.org/10.1016/0166-218X(95)00095-9}
  {\path{doi:10.1016/0166-218X(95)00095-9}}.

\bibitem{MonienS88}
B.~Monien and I.~H. Sudborough.
\newblock Min cut is {NP}-complete for edge weighted treees.
\newblock {\em Theor. Comput. Sci.}, 58(1-3):209--229, 1988.
\newblock \href {http://dx.doi.org/10.1016/0304-3975(88)90028-X}
  {\path{doi:10.1016/0304-3975(88)90028-X}}.

\bibitem{Niedermeier10}
R.~Niedermeier.
\newblock Reflections on multivariate algorithmics and problem
  parameterization.
\newblock In {\em Proc. STACS 2010}, volume~5 of {\em LIPIcs}, pages 17--32,
  2010.
\newblock \href {http://dx.doi.org/10.4230/LIPIcs.STACS.2010.2495}
  {\path{doi:10.4230/LIPIcs.STACS.2010.2495}}.

\bibitem{OlesenM02}
K.~G. Olesen and A.~L. Madsen.
\newblock Maximal prime subgraph decomposition of {B}ayesian networks.
\newblock {\em IEEE T. Syst. Man Cy. B}, 32(1):21--31, 2002.
\newblock \href {http://dx.doi.org/10.1109/3477.979956}
  {\path{doi:10.1109/3477.979956}}.

\end{thebibliography}
\bibliographystyle{abbrvurl}

% Images that do not fit in the main text, and do not add much extra value to the appendix. 
% TODO: Put these back somewhere for the journal paper?
%
%\begin{figure}[htb]
%\begin{center}
%\includegraphics[scale=0.9]{simplicial}
%\end{center}
%\caption{(a) $v$ is a simplicial vertex. (b) $v$ is an almost simplicial
%vertex with special neighbor $w$. Dotted lines can be edges or pairs of
%nonadjacent vertices.}
%\label{figure:simplicial}
%\end{figure}

%\begin{figure}[htb]
%\begin{center}
%\includegraphics[scale=0.9]{commonneighbors}
%\end{center}
%\caption{An example of an application of the Common neighbors improvement
%rule. $k\leq 3$.}
%\label{figure:commonneighbors}
%\end{figure}

\end{document}